\documentclass[12pt,letterpaper]{article}
\usepackage{amssymb}

\usepackage{algorithm}
\usepackage{algorithmic}
\usepackage{cite}
\usepackage{graphicx}
\usepackage{subfigure}
\usepackage{amsmath}
\usepackage{amssymb}
\usepackage{latexsym}
\usepackage{empheq}
\usepackage{dsfont}
\usepackage{hyperref}

\newcommand{\trace}[1]{\mbox{Tr}\left({#1}\right)}
\newcommand{\vvec}[1]{\mbox{vec}\left({#1}\right)}
\newcommand{\conj}[1]{\mbox{conj}\left({#1}\right)}
\newcommand{\pr}[1]{\mbox{Pr}\left\{{#1}\right\}}
\newcommand{\re}[1]{\mbox{Re}\left\{{#1}\right\}}

\newcommand{\ud}{\mathrm{d}}

\newtheorem{theorem}{Theorem}

\newtheorem{coro}{Corollary}

\newtheorem{remark}{Remark}
\newtheorem{crit}{Criterion}

\begin{document}

\title{Hybrid ARQ in Multiple-Antenna Slow Fading Channels: Performance Limits and Optimal Linear Dispersion Code Design}
\author{\normalsize
Cong Shen
\and
\normalsize Michael P. Fitz \footnote{The material in this paper was
presented in part at the 42nd Annual Asilomar Conference on Signals,
Systems, and Computers, Pacific Grove, CA, USA, Oct. 2008.}
\footnote{Cong Shen is with the Department of Electrical
Engineering, University of California Los Angeles (UCLA), Los
Angeles, CA 90095, USA. Email: \texttt{congshen@ee.ucla.edu}.
Michael P. Fitz was with the Department of Electrical Engineering,
University of California Los Angeles (UCLA), Los Angeles, CA 90095,
USA. He is now with Northrop Grumman Space Technology, Redondo
Beach, CA 90278, USA. Email: \texttt{Michael.Fitz@ngc.com}.}}

\date{\today}

\maketitle

\begin{abstract}
This paper focuses on studying the fundamental performance limits
and linear dispersion code design for the MIMO-ARQ slow fading
channel. Optimal average rate of well-known HARQ protocols is
analyzed. The optimal design of space-time coding for the MIMO-ARQ
channel is discussed. Information-theoretic measures are used to
optimize the rate assignment and derive the optimum design
criterion, which is then used to evaluate the optimality of existing
space-time codes. A different design criterion, which is obtained
from the error probability analysis of space-time coded MIMO-HARQ,
is presented. Examples are studied to reveal the gain of ARQ
feedback in space-time coded MIMO systems.
\end{abstract}


\section{Introduction}
\label{sec:intro}

Hybrid ARQ (HARQ) techniques combine the automatic-repeat-request
(ARQ) feedback with the forward error correction (FEC) codes to
achieve better reliability and higher throughput \cite{LCM:84}. They
are widely used in most of the contemporary communication systems,
as the delay tolerance of many data services allows using
retransmission to recover erroneous packets. A good summary of the
progress of HARQ schemes is presented in \cite{CHIW:98}. From an
information-theoretic perspective, HARQ systems can be viewed as
\emph{channel with sequential feedback}, and there has been some
research on the fundamental limits of this channel. An
information-theoretic throughput and delay analysis of several HARQ
schemes in a Gaussian collision channel is presented in
\cite{CT:01}. Throughput analysis of incremental redundancy HARQ in
the block-fading additive white Gaussian noise (AWGN) channel is
carried out in \cite{SCV:04}. Optimal average rate performance of a
scalar slow fading channel with ARQ feedback is done in
\cite{SLF2:08,SLF:08}.

Most of the aforementioned research is restricted to scalar
channels. A natural question is how to incorporate the design of
HARQ into multiple-input-multiple-output (MIMO) systems, and it has
sparked many research activities. Most existing literature regarding
MIMO-HARQ design falls into the following two categories. One is the
information-theoretic analysis, for which the focus is on extending
the Diversity-Multiplexing Tradeoff (DMT) framework \cite{ZT:03} to
consider ARQ feedback \cite{GCD:06,KS:07it,CGRC:08}. The other is
the joint design of MIMO transmission and ARQ feedback. Receiver
processing of ARQ retransmissions is studied in
\cite{ODHJ:03,OKP:04}. It has been shown that multiple ARQ packets
combining is non-trivial in a MIMO setting, especially when
soft-output is desired \cite{JLLC:07}. For the transmitter design,
two different approaches have been taken to exploit the additional
spatial degrees of freedom. One is to study the bits/symbols
rearrangement for retransmissions to exploit diversity
\cite{ZLH:02,KMY:04,MPGO:07,CP:08}, and the other is to explore
linear precoder design \cite{NI:01,ZLWW:07}.


Despite the many efforts in studying MIMO-HARQ design, there are
still some important problems left unanswered. From the perspective
of information-theoretic  analysis, the DMT-based approach focuses
only on the high Signal-to-Noise Ratio (SNR) asymptotics and gives
the tradeoff between multiplexing gain (pre-log), diversity gain
(error exponent) and ARQ delay. It also requires a family of
space-time codes (STC) whose rates scale with the SNR. In practice,
however, one is also interested in designing a fixed-rate STC that
operates well within a range of finite SNRs. It is unclear what is
the optimal MIMO-HARQ throughput in this scenario, and how to design
the optimal STC to achieve it. For the practical HARQ retransmission
protocol design, the aforementioned works provide separate and ad
hoc designs. There is no general framework that unifies the existing
results and guides the optimal design.

This paper is devoted to solving these problems. We begin by
introducing the system model in Sec.~\ref{sec:model}. The
performance limits of existing HARQ protocols in a MIMO setting are
investigated in Sec.~\ref{sec:existingHARQ}. Sec.~\ref{sec:LDC_MI}
studies the fundamental average rate limit of STC-based MIMO-HARQ
protocols, and gives the optimal STC design criterion. Many existing
STCs are re-visited under this framework, both analytically and
numerically. A different approach based on the error probability
analysis is given in Sec.~\ref{sec:LDC_error}, where the so-called
\emph{$n$-th pairwise error probability} plays a key role in the
analysis. Optimal STC design criterion based on the error
probability analysis is proposed and exemplary STCs are studied.
Finally, Sec.~\ref{sec:conc} concludes the paper.

Throughout the paper the following notations will be used. Matrices
and vectors are denoted with bold capital and lowercase letters,
respectively. $\mathbf{A}(i,j)$ is the $(i,j)$-th element of the
matrix $\mathbf{A}$, and $\mathbf{a}(i)$ is the $i$-th element of
the vector $\mathbf{a}$. $\mathbb{E}[\cdot]$ denotes the expectation
of a random variable. $a^*$ denotes the conjugate of the complex
number $a$, $\re{a}$ is the real part of $a$, and $\mathbf{a}^t$ is
the transpose of vector $\mathbf{a}$. We use $\trace{\mathbf{A}}$
and $||\mathbf{A}||_F$ to denote the trace and Frobenius norm of
matrix $\mathbf{A}$, respectively. $\mathbf{A}^H$ and
$\conj{\mathbf{A}}$ are the Hermitian and the conjugate of the
complex matrix $\mathbf{A}$, respectively. The Kronecker product
operation of two matrices is denoted by $\otimes$, and
$\vvec{\cdot}$ is the vectorization operation. Finally,
$|\mathcal{W}|$ denotes the cardinality of the set $\mathcal{W}$.

\section{System model}
\label{sec:model} In this section, the MIMO channel model with ARQ
feedback is described, and existing HARQ protocols that will be
analyzed in Sec.~\ref{sec:existingHARQ} are introduced.

\subsection{MIMO channel with ARQ feedback}
\label{sec:model_CH}

This paper studies a single-user multiple-antenna wireless system
equipped with $L_t$ transmit and $L_r$ receive antennas. It is
assumed that there is an error-free and delay-free ARQ feedback
link, which indicates to the transmitter the successful decoding by
acknowledgement (ACK) and failed decoding by negative
acknowledgement (NACK). The maximum number of ARQ transmissions is
denoted by $N$, which is also called the maximum allowable ARQ
rounds in the literature. With this constraint, if after a total of
$N$ transmissions the receiver still cannot decode the message
successfully, no further attempt will be tried and a decoding
failure is declared.

It is assumed there is a very large pool of information messages
available at the transmitter. As soon as the transmission of current
message (including possible retransmissions) is done, the next
message is encoded and transmitted immediately. Consider a set of
uniformly distributed messages $\mathcal{W}$. For the transmission
of each message, the information message $w \in \mathcal{W}$ is sent
to a MIMO-HARQ encoder, which generates $N$ matrix sub-codewords
$\left\{ \mathbf{X}_n(w) \right\}_{n=1}^{N}$, each corresponds to
one ARQ round. Sub-codeword $\mathbf{X}_n(w)$ will be transmitted in
the $n$-th ARQ round if the decoding fails at the $(n-1)$-th round.
Denote $\mathbf{X}^{(n)}(w) \doteq \left[ \mathbf{X}_1(w), \cdots,
\mathbf{X}_n(w) \right]$, $\forall n=1, \cdots, N$, which is the
overall codeword after the $n$-th ARQ round. The receiver performs
optimum decoding based on all the received packets in ARQ round $1,
\cdots, n$. Each sub-codeword satisfies $\mathbf{X}_n(w) \in
\mathcal{C}^{L_t \times L_n/r_n}$, where $L_n/r_n$ denotes the
length of the sub-codeword, $r_n$ is determined by the ``coding
rate'' of the space-time structure as $r_n \doteq \frac{K_n}{T_n}$
where $K_n$ and $T_n$ are the number of symbols and the time slots
per space-time codeword, respectively, and $L_n$ is the
capacity-achieving channel code length. With this model, the overall
transmission rate in the $n$-th ARQ round is
\begin{equation}
\label{eqn:ovral_rate}
R^{(n)} = \frac{\log_2{|\mathcal{W}|}}{\sum_{i=1}^{n} L_i/r_i}, \forall n=1, \cdots, N.
\end{equation}

The MIMO channel $\mathbf{H} \in \mathcal{C}^{L_r \times L_t}$ is
assumed to be a random matrix with independent and identically
distributed (i.i.d.) complex circularly symmetric Gaussian entries
with unit variance. The channel matrix $\mathbf{H}$ is assumed to be
constant within the channel coherence time ($T_c$ symbols), and
changes independently to a different value in the next coherence
block. Depending on the relationship between $T_c$ and $\left\{
L_n/r_n \right\}_{n=1}^{N}$, different MIMO-HARQ models have been
studied in the literature \cite{GCD:06,CGRC:08}. In this work, a
slow fading MIMO channel model is considered, where $T_c \gg
\sum_{n=1}^{N}L_n/r_n$. This means the MIMO channel remains
unchanged throughout all possible retransmissions. This models the
worst-case scenario compared to \cite{GCD:06,CGRC:08}, as there is
no \emph{time diversity} to exploit by retransmissions.

\begin{figure}
\centering
\includegraphics[width=\textwidth]{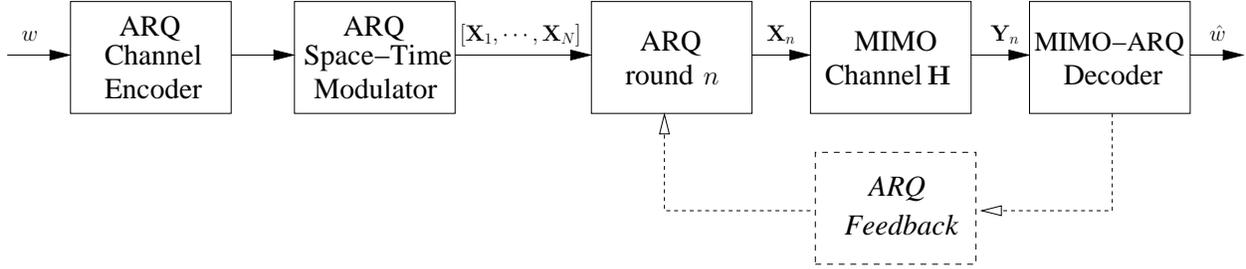}
\caption{The MIMO-HARQ system model.} \label{fig:sys_model}
\end{figure}

With the described model, the received signal corresponding to the $n$-th transmission can be written as
\begin{equation}
\mathbf{Y}_n  = \sqrt{\frac{\textrm{$\sf{SNR}$}}{L_t}}\mathbf{H}\mathbf{X}_n+\mathbf{Z}_n
\end{equation}
and the overall received signal after the $n$-th ARQ round is
\begin{equation}
\mathbf{Y}^{(n)}  = \sqrt{\frac{\textrm{$\sf{SNR}$}}{L_t}}\mathbf{H}\mathbf{X}^{(n)}+\mathbf{Z}^{(n)},
\end{equation}
where $\mathbf{Y}^{(n)} \doteq \left[ \mathbf{Y}_1, \cdots,
\mathbf{Y}_n \right]$ and $\mathbf{X}^{(n)}$, $\mathbf{Z}^{(n)}$ are
similarly defined; the additive noise $\mathbf{Z}^{(n)}$ has i.i.d.
entries $\mathbf{Z}^{(n)}(i,j) \sim \mathcal{CN}(0,1)$. The overall
transmit codewords $\mathbf{X}^{(N)}(w)$ are normalized to satisfy
the average power constraint\footnote{The transmit power constraint
will be discussed in more detail when the optimal LDC design for
MIMO-HARQ is presented in Sec.~\ref{sec:LDC_MI}.}
\begin{equation}
\frac{1}{\sum_{n=1}^{N} L_n/r_n} \mathbb{E}||\mathbf{X}^{(N)}(w)||_F^2 \leq L_t
\end{equation}
where the expectation is over all possible codewords. Therefore, the
average SNR per receive antenna is $\sf{SNR}$. The receiver is
assumed to have perfect knowledge of $\mathbf{H}$. This is a
reasonable assumption since the slowly varying nature of the channel
helps receiver channel estimation. The transmitter is assumed to
have no knowledge of $\mathbf{H}$ prior to the transmission, but it
should be noted that ARQ retransmissions essentially inform partial
CSI to the transmitter in a sequential fashion \cite{SLF:08}.
Fig.~\ref{fig:sys_model} illustrates the system model under
consideration.

\subsection{Hybrid ARQ strategies}
\label{sec:model_ARQ}



HARQ incorporates FEC coding into ARQ feedback to enhance the
retransmission performance. Depending on transmitter's
retransmission strategy and receiver's decoding from multiple
packets, several efficient protocols have been proposed and studied.
In this work, two traditional stop-and-wait HARQ protocols are
considered:
\begin{itemize}
\item [1)] \textbf{Incremental Redundancy (IR)}. This is a
\emph{code combining} scheme \cite{Wicker:95}. At the transmitter,
information message is encoded into an overall codeword of length
$\sum_{n=1}^{N} L_n$, and then the codeword is \emph{serially}
punctured into $N$ sub-codewords each with length $L_n$. These
sub-codewords are mapped into the transmit matrix codewords $\left[
\mathbf{X}_1, \cdots, \mathbf{X}_N \right]$. At the $n$-th
transmission, $\mathbf{X}_n$ is sent through the channel as
additional (when $n>1$) redundancy symbols to reduce the
transmission rate. The receiver tries to decode the message based on
all the packets it receives up until round $n$.
\item [2)] \textbf{Chase Combining (CC)}. This belongs to the
category of \emph{diversity combining} \cite{Wicker:95}. Upon each
retransmission request, the transmitter simply repeats the same
packet: $\mathbf{X}_n = \mathbf{X}_1$, $\forall n=2, \cdots, N$.
Different diversity combining schemes can be used at the receiver,
among which the Chase combining \cite{Chase:85}, which essentially
is a maximum ratio combining of all the received packets in the
scalar channel, gives the best performance.
\end{itemize}
These well-known HARQ protocols will be analyzed in a MIMO setting
and the optimal performance is reported in
Section~\ref{sec:existingHARQ}.


Traditional HARQ protocols are designed to extract \emph{time}
diversity in a scalar fading channel. The performance depends
critically on the channel variation over retransmissions. The slow
fading channel model considered in this work allows no time
diversity for the HARQ protocols to exploit, and hence
packet-combining-based schemes have limited performance
\cite{SLF:08}. Multiple antenna system provides additional
dimensions for \emph{spatial} diversity, which is not effectively
exploited by traditional protocols. On the other hand, although
provides the best performance, IR has higher complexity due to the
generation of rate-compatible channel codes or the rateless codes.
Both problems call for the design of HARQ protocols that jointly
consider the space-time structure to exploit spatial diversity and
low-complexity ARQ retransmissions. This is the focus of
Section~\ref{sec:LDC_MI} and \ref{sec:LDC_error}.


The figure of merit in this paper is the \emph{long-term throughput}, and an information-theoretic view is taken for the MIMO-HARQ design problem. The throughput of a HARQ protocol is defined \cite{Wicker:95} as the average number of bits accepted by the receiver in the time it takes to send the data packet. As discussed in \cite{SLF:08}, how to compute the throughput depends on the specific applications and the assumptions on how to use the channel. Typical derivation involves the evaluation of average transmission time. \emph{Renewal theory} has been successfully applied to this problem \cite{Zorzi:96,CT:01}. However, this formulation leads to the problem that \emph{the empirical channel distribution does not match the true channel statistics}, and the reason is that the length of channel uses is determined by the instantaneous channel state. This problem has been discussed in \cite{WJ:08,SLF:08}, and it has been shown \cite{SLF:08} that for some applications, \emph{average rate} \cite{LLTF:02,SS:03,EGL:07} gives the long-term successful communication rate and is a more reasonable metric. The average rate is chosen as the performance measure in this work.

\section{Performance limits of existing HARQ protocols}
\label{sec:existingHARQ}

This section studies the optimal average rate performance where the existing HARQ protocols are directly applied in a vector channel. We will first derive the average rate for general HARQ systems, and then apply the result to IR and CC. The general average rate expression is a fundamental result and will also be useful in developing optimal space-time HARQ in Section~\ref{sec:LDC_MI}.

Before analyzing the average rate, some important events are defined and the associated probabilities are studied. Due to the randomness of the channel matrix $\mathbf{H}$ and the limit $N$ on the ARQ deadline, the successful communication rate $R$ is a random variable. Define two events for ARQ round $n$, $\forall n=1, \cdots, N$:
\begin{eqnarray}
\mathcal{A}_n &=&  \textrm{Decoding at the end of ARQ round $n$ is successful}; \label{def:evnt_An} \\
\mathcal{S}_n &=&  \textrm{ARQ round $n$ is activated}  \label{def:evnt_Sn}.
\end{eqnarray}
There are two important observations about these events, both originated from an information theoretic viewpoint\footnote{As one will see in Sec.~\ref{sec:LDC_error}, finite coding length leads to some different conclusions for the HARQ design.}. First, \emph{a decoding failure is equivalent to a channel outage}, i.e., the instantaneous channel capacity cannot support the current transmission rate. This is because long-blocklength capacity-achieving channel coding ``averages out'' the additive noise. Thus one has
\begin{equation}
\label{eqn:eventAn}
\pr{ \mathcal{A}_n } = \pr{ C^{(n)}\left( \mathbf{H}_{\textrm{$\sf{eq}$}}^{(n)} \right) \geq R^{(n)}}.
\end{equation}
where $C^{(n)}( \mathbf{H}_{\textrm{$\sf{eq}$}}^{(n)} )$ is the equivalent channel capacity at ARQ round $n$;  $R^{(n)}$ is the overall communication rate at round $n$, which is given in Equation~(\ref{eqn:ovral_rate}). The second observation is that \emph{additional transmissions can only increase the successful decoding probability}, via either increasing $C^{(n)}( \mathbf{H}_{\textrm{$\sf{eq}$}}^{(n)} )$ or reducing $R^{(n)}$, or both. This observation suggests
\begin{equation}
\label{eqn:eventAn_subset}
\mathcal{A}_{n-1} \subseteq \mathcal{A}_{n},
\end{equation}
and hence
\begin{eqnarray}
\pr{ \mathcal{S}_n } &=& \pr{ \overline{\mathcal{A}_1}, \cdots, \overline{\mathcal{A}_{n-1}} } \nonumber \\
&=& \pr{ \overline{\mathcal{A}_{n-1}} } \label{eqn:eventSn}.
\end{eqnarray}
These observations play an important role in analyzing the information-theoretic performance limits of MIMO-HARQ.

With the definitions (\ref{def:evnt_An}) and (\ref{def:evnt_Sn}), the successful communication rate $R$ can be expressed as
\begin{eqnarray}
R &=& \left\{ \begin{array}{ll}
R^{(n)}, & \textrm{if $\left\{ \mathcal{S}_n, \mathcal{A}_n \right\}$}, \forall n=1, \cdots, N \\
0, & \textrm{if $\overline{\mathcal{A}_N}$}
\end{array} \right.   \\
&=& \left\{ \begin{array}{ll}
R^{(n)}, & \textrm{if $\left\{ \overline{\mathcal{A}_{n-1}}, \mathcal{A}_n \right\}$}, \forall n=1, \cdots, N \\
0, & \textrm{if $\overline{\mathcal{A}_N}$}
\end{array} \right. \label{eqn:R}
\end{eqnarray}
where the second equality comes from (\ref{eqn:eventSn}), and the convention $\pr{ \mathcal{A}_{0} } = 0$ is used. The average rate is then
\begin{eqnarray}
\bar{R} &=& \sum_{n=1}^{N} R^{(n)} \pr{\overline{\mathcal{A}_{n-1}}, \mathcal{A}_n} \label{eqn:avgrate0} \\
&=& \sum_{n=1}^{N} R^{(n)} \pr{ C^{(n-1)}\left( \mathbf{H}_{\textrm{$\sf{eq}$}}^{(n-1)} \right) < R^{(n-1)}, C^{(n)}\left( \mathbf{H}_{\textrm{$\sf{eq}$}}^{(n)} \right) \geq R^{(n)}}, \label{eqn:avgrate1}
\end{eqnarray}
where we define $R^{(0)}=\infty$ (since the denominator in (\ref{eqn:ovral_rate}) is zero) and $C^{(0)}\left( \mathbf{H}_{\textrm{$\sf{eq}$}} \right) = 0$. A different expression can be obtained if one notices that
\begin{equation}
\pr{\overline{\mathcal{A}_{n-1}}, \mathcal{A}_n} = \pr{ \mathcal{A}_n } - \pr{ \mathcal{A}_{n-1} },
\end{equation}
which is true since $\mathcal{A}_{n-1} \subseteq \mathcal{A}_{n}$. The average rate expression (\ref{eqn:avgrate0}) then can be written as
\begin{eqnarray}
\bar{R} &=& \sum_{n=1}^{N} R^{(n)} \left( \pr{ \mathcal{A}_n } - \pr{ \mathcal{A}_{n-1} } \right) \nonumber  \\
&=& \sum_{n=1}^{N} \left( R^{(n)} - R^{(n+1)} \right)  \pr{ \mathcal{A}_n } \label{eqn:avgrate20} \\
&=& \sum_{n=1}^{N} \left( R^{(n)} - R^{(n+1)} \right)  \pr{ C^{(n)}\left( \mathbf{H}_{\textrm{$\sf{eq}$}}^{(n)} \right) \geq R^{(n)} }, \label{eqn:avgrate2}
\end{eqnarray}
where we define $R^{(N+1)}=0$ (since the retransmission stops after the $n$-th round). It should be noted that the term $R^{(n)} - R^{(n+1)}$ denotes the rate decrease in the $(n+1)$-th transmission.

The equivalent channel capacity $C^{(n)}( \mathbf{H}_{\textrm{$\sf{eq}$}}^{(n)} )$ at ARQ round $n$ is determined not only by the instantaneous channel matrix $\mathbf{H}$, but also by the specific HARQ protocol and the transmit covariance matrices $\left\{\mathbf{K}_{i}\right\}_{i=1}^{n}$, where $\mathbf{K}_i$ is the transmit covariance matrix at ARQ round $i$ and could be different over retransmissions. As one shall see in the sequel, these degrees of freedom shall be carefully exploited in the MIMO-HARQ protocol design.

With the general average rate expressions (\ref{eqn:avgrate1}) and (\ref{eqn:avgrate2}), we proceed to study IR and CC, respectively.

\subsection{Incremental Redundancy}


The average rate for general IR can be obtained directly from (\ref{eqn:avgrate2}):
\begin{equation}
\label{eqn:avgrateIRgen}
\bar{R}_{\textrm{$\sf{ir}$}} = \sum_{n=1}^{N} \left( R^{(n)} - R^{(n+1)} \right)  \pr{ C_{\textrm{$\sf{ir}$}}^{(n)}\left( \mathbf{H}_{\textrm{$\sf{eq, ir}$}}^{(n)} \right) \geq R^{(n)} },
\end{equation}
and the remaining problem is to analyze the capacity of IR after each ARQ round. The capacity for a given MIMO channel $\mathbf{H}$ and total transmit covariance matrix  $\mathbf{K}$ is known to be \cite{Tel:95}
\begin{equation}
\label{eqn:MIMOcap}
C_{\textrm{$\sf{mimo}$}}\left( \mathbf{H}, \mathbf{K} \right) = \log \det \left( \mathbf{I}_{L_r} + \mathbf{H} \mathbf{K} \mathbf{H}^{H} \right)
\end{equation}
where $\mathbf{K}$ has to satisfy the power constraint\footnote{The constraint (\ref{eqn:MIMOcap_PC}) assumes that each ARQ round consumes the same average power. One can also dynamically allocate power among ARQ rounds, which leads to a different power constraint and average rate expression. The case for scalar fading channel has been addressed in \cite{SLF:08}.}
\begin{equation}
\label{eqn:MIMOcap_PC}
\trace{\mathbf{K}} \leq \textrm{$\sf{SNR}$}.
\end{equation}
Assuming transmit covariance matrix $\mathbf{K}_n$ is used in ARQ round $n$, the overall IR capacity where different portions have different covariance matrices is a TDMA-type one:
\begin{eqnarray}
C_{\textrm{$\sf{ir}$}}^{(n)}\left( \mathbf{H}_{\textrm{$\sf{eq, ir}$}}^{(n)} \right) &=& \frac{ \sum_{i=1}^{n} C_{\textrm{$\sf{mimo}$}} \left( \mathbf{H}, \mathbf{K}_i \right)L_i/r_i }{ \sum_{i=1}^{n} L_i/r_i } \nonumber \\
&=&  \sum_{i=1}^{n} \left( \frac{R^{(n)}}{R^{(i)}} - \frac{R^{(n)}}{R^{(i-1)}} \right) C_{\textrm{$\sf{mimo}$}}\left( \mathbf{H}, \mathbf{K}_i \right) \label{eqn:avgrateIRgen1}.
\end{eqnarray}
where (\ref{eqn:avgrateIRgen1}) is obtained from (\ref{eqn:ovral_rate}).

The general IR average rate maximization problem can now be formulated as
\begin{equation}
\label{prb:IRavgrate_gen}
\begin{array}{ll}
\underset{\left\{  \mathbf{K}_n, R_n \right\}_{n=1}^{N}}{\mbox{maximize}} & \sum_{n=1}^{N} \left( R^{(n)} - R^{(n+1)} \right)  \pr{ \sum_{i=1}^{n} \left( \frac{R^{(n)}}{R^{(i)}} - \frac{R^{(n)}}{R^{(i-1)}} \right) C_{\textrm{$\sf{mimo}$}}\left( \mathbf{H}, \mathbf{K}_i \right)  \geq R^{(n)} } \\
\mbox{subject to } & R^{(N)} \geq 0 \\
{} & R^{(n)} \leq R^{(n-1)}, \forall n=1, \cdots, N \\
{} & \trace{\mathbf{K}_n} \leq \textrm{$\sf{SNR}$}, \forall n=1, \cdots, N
\end{array}
\end{equation}
Solving this problem is extremely difficult, which comes mainly from the search of optimal covariance matrices $\left\{  \mathbf{K}_n \right\}_{n=1}^{N}$. Evaluating the objective function in Problem (\ref{prb:IRavgrate_gen}) requires  the distribution of MIMO capacity $C_{\textrm{$\sf{mimo}$}}\left( \mathbf{H}, \mathbf{K} \right)$ as a function of the covariance matrix $ \mathbf{K}$, which is an unsolved problem. In fact, even the simpler problem of finding the optimal covariance matrix that minimizes the outage probability is still open.

To make progress, this paper will focus on the isotropic Gaussian input distribution over all transmit antennas:
\begin{equation}
\label{eqn:MIMOisotr}
\mathbf{K}_n =
\mathbf{K}=\frac{\textrm{$\sf{SNR}$}}{L_t} \mathbf{I}_{L_t}.
\end{equation}
This is by no means the optimal solution but very easy to deal with and widely assumed in literature. With (\ref{eqn:MIMOisotr}) the IR capacity is the mutual information of a MIMO channel with isotropic Gaussian input distribution:
\begin{eqnarray}
C_{\textrm{$\sf{ir}$}}^{(n)}\left( \mathbf{H}_{\textrm{$\sf{eq, ir}$}}^{(n)} \right) &=& C_{\textrm{$\sf{mimo}$}}\left( \mathbf{H}, \frac{\textrm{$\sf{SNR}$}}{L_t} \mathbf{I}_{L_t} \right)  \nonumber \\
&=&  \log \det \left( \mathbf{I}_{L_r} +  \frac{\textrm{$\sf{SNR}$}}{L_t} \mathbf{H} \mathbf{H}^{H} \right) \label{eqn:MIMOcap_isotr},
\end{eqnarray}
and the IR average rate maximization problem becomes
\begin{equation}
\label{prb:IRavgrate_isotr}
\begin{array}{ll}
\underset{\left\{ R_n \right\}_{n=1}^{N}}{\mbox{maximize}} & \sum_{n=1}^{N} \left( R^{(n)} - R^{(n+1)} \right)  \pr{ C_{\textrm{$\sf{mimo}$}}\left( \mathbf{H}, \frac{\textrm{$\sf{SNR}$}}{L_t} \mathbf{I}_{L_t} \right)  \geq R^{(n)} } \\
\mbox{subject to } & R^{(N)} \geq 0 \\
{} & R^{(n)} \leq R^{(n-1)}, \forall n=1, \cdots, N.
\end{array}
\end{equation}
For the remaining of this paper, the isotropic Gaussian input distribution is assumed unless the covariance matrix is explicitly specified.

\subsection{Chase Combining}
In CC, each retransmission effectively increases the number of receive antennas by $L_r$ via the repetition of previous packet. Thus after ARQ round $n$, the equivalent system is a $(L_t, nL_r)$ MIMO with an equivalent channel matrix
\begin{equation}
\mathbf{H}_{\textrm{$\sf{eq, cc}$}}^{(n)} = \mathds{1}_n \otimes  \mathbf{H},
\end{equation}
and
\begin{equation}
\label{eqn:CCcap_gen}
C_{\textrm{$\sf{cc}$}}^{(n)}\left( \mathbf{H}_{\textrm{$\sf{eq, cc}$}}^{(n)} \right) = \frac{1}{n} C_{\textrm{$\sf{mimo}$}}\left( \mathbf{H}_{\textrm{$\sf{eq, cc}$}}^{(n)}, \mathbf{K} \right).
\end{equation}
Meanwhile, since each retransmission is a simple packet repetition, the effective rate after ARQ round $n$ is
\begin{equation}
R^{(n)} = \frac{R}{n},
\end{equation}
where $R$ is the rate in ARQ round 1. Hence, the general CC average rate maximization problem is formulated as
\begin{equation}
\label{prb:CCavgrate_gen}
\begin{array}{ll}
\underset{ \mathbf{K}, R }{\mbox{maximize}} & \sum_{n=1}^{N} \frac{R}{n} \pr{ C_{\textrm{$\sf{mimo}$}}\left( \mathbf{H}_{\textrm{$\sf{eq, cc}$}}^{(n-1)}, \mathbf{K} \right) < R,  C_{\textrm{$\sf{mimo}$}}\left( \mathbf{H}_{\textrm{$\sf{eq, cc}$}}^{(n)}, \mathbf{K} \right) \geq R } \\
\mbox{subject to } & R \geq 0 \\
{} & \trace{\mathbf{K}} \leq \textrm{$\sf{SNR}$}.
\end{array}
\end{equation}
Note that due to the simple packet repetition in retransmissions, only one covariance matrix $\mathbf{K}$ and one rate parameter $R$ need to be optimized. If the isotropic Gaussian input distribution is assumed, the problem becomes ($R \geq 0$ can be dropped):
\begin{equation}
\label{prb:CCavgrate_isotr}
\underset{ R }{\mbox{maximize}}  \sum_{n=1}^{N} \frac{R}{n} \pr{ C_{\textrm{$\sf{mimo}$}}\left( \mathbf{H}_{\textrm{$\sf{eq, cc}$}}^{(n-1)},  \frac{\textrm{$\sf{SNR}$}}{L_t} \mathbf{I}_{L_t} \right) < R,  C_{\textrm{$\sf{mimo}$}}\left( \mathbf{H}_{\textrm{$\sf{eq, cc}$}}^{(n)},  \frac{\textrm{$\sf{SNR}$}}{L_t} \mathbf{I}_{L_t} \right) \geq R }.
\end{equation}
Further simplification is possible if one notices that
\begin{eqnarray}
(\mathbf{H}_{\textrm{$\sf{eq, cc}$}}^{(n)}) (\mathbf{H}_{\textrm{$\sf{eq, cc}$}}^{(n)})^{H} &=& (\mathds{1}_n \otimes  \mathbf{H}) (\mathds{1}_n^{T} \otimes  \mathbf{H}^{H}) \nonumber \\
&=& \mathds{1}_n \mathds{1}_n^{T} \otimes \mathbf{H}\mathbf{H}^{H},
\end{eqnarray}
and
\begin{equation}
C_{\textrm{$\sf{mimo}$}}\left( \mathbf{H}_{\textrm{$\sf{eq, cc}$}}^{(n)},  \frac{\textrm{$\sf{SNR}$}}{L_t} \mathbf{I}_{L_t} \right) = \log \det \left( \mathbf{I}_{nL_r} + \frac{\textrm{$\sf{SNR}$}}{L_t} \mathds{1}_n \mathds{1}_n^{T} \otimes \mathbf{H}\mathbf{H}^{H}\right).
\end{equation}

%

\subsection{Examples}
To evaluate the performance of IR and CC, several exemplary MIMO configurations are considered and the corresponding optimal average rate performance (\ref{prb:IRavgrate_isotr}) (\ref{prb:CCavgrate_isotr}) is compared in this section.

We start with the case of $L_r=1$ (MISO)\footnote{The case of SIMO ($L_t=1$) can be similarly analyzed.}, which is easy to analyze as the received signal expands only one spatial dimension, and hence the capacity distribution would be easy to obtain. For a $(L_t, 1)$ MISO system, the capacity (\ref{eqn:MIMOcap_isotr}) becomes
\begin{equation}
\label{eqn:MISOcap_isotr}
C_{\textrm{$\sf{mimo}$}}\left(
\mathbf{h}^{T},  \frac{\textrm{$\sf{SNR}$}}{L_t} \mathbf{I}_{L_t}
\right) = \log \left( 1 +  \frac{\textrm{$\sf{SNR}$}}{L_t}
\sum_{i=1}^{L_t} |h_i|^2 \right).
\end{equation}
Define
\begin{equation}
\label{eqn:def_g_Lt}
g_{L_t} \doteq \sum_{i=1}^{L_t} |h_i|^2,
\end{equation}
the distribution of $C_{\textrm{$\sf{mimo}$}}\left( \mathbf{h}^{T},  \frac{\textrm{$\sf{SNR}$}}{L_t} \mathbf{I}_{L_t} \right)$ is then completely characterized by the random variable $g_{L_t}$. If the attention is restricted to a Gaussian channel with $h_i \sim \mathcal{CN}(0,1)$, then $g_{L_t}$ follows a Chi-square distribution with $2 L_t$ degrees of freedom and it has a PDF
\begin{equation}
\label{eqn:MISOPDFChi}
f(g) = \frac{1}{\Gamma\left( L_t \right)} g^{L_t-1} e^{-g}, \qquad g \geq 0,
\end{equation}
where $\Gamma\left( L_t \right) = \int_{0}^{\infty} t^{L_t-1} e^{-t} \ud t$ is the Gamma function. The CDF of $g_{L_t}$,  $F_G(g) \doteq \pr{ g_{L_t} \leq g} = \int_{0}^{\infty} f(t) \ud t$, can also be calculated and a closed-form expression is given in \cite{Simon:02}. The CDF of the MISO capacity then reduces to
\begin{equation}
\label{eqn:MISOcapCDFequv}
\pr {C_{\textrm{$\sf{mimo}$}}\left( \mathbf{h}^{T},  \frac{\textrm{$\sf{SNR}$}}{L_t} \mathbf{I}_{L_t} \right) \leq R } = \pr{ g_{L_t} \leq \frac{(2^{R}-1) L_t}{\textrm{$\sf{SNR}$}} } = F_G\left( \frac{(2^{R}-1) L_t}{\textrm{$\sf{SNR}$}} \right).
\end{equation}

The equivalent IR channel capacity is given in (\ref{eqn:MISOcap_isotr}), and the average rate with isotropic Gaussian input distribution for IR can now be re-written as
\begin{equation}
\bar{R}_{\textrm{$\sf{ir}$}} =  \sum_{n=1}^{N} \left( R^{(n)} - R^{(n+1)} \right)  \left( 1 - F_G\left(  \frac{(2^{R}-1) L_t}{\textrm{$\sf{SNR}$}} \right) \right).
\end{equation}
As for CC, the interesting observation is that the impact of each retransmission on the overall capacity is only to increase the receive SNR to $\sf{SNR}$$_n = n \sf{SNR}$. Thus,
\begin{equation}
C_{\textrm{$\sf{cc}$}}^{(n)}\left( \mathbf{H}_{\textrm{$\sf{eq, cc}$}}^{(n)} \right) = \frac{1}{n} \log \left( 1 + \frac{n \textrm{$\sf{SNR}$}}{L_t} g_{L_t} \right),
\end{equation}
and the average rate becomes
\begin{equation}
\bar{R}_{\textrm{$\sf{cc}$}} = \sum_{n=1}^{N} \frac{R}{n} \left( F_G\left( \frac{(2^{R}-1) L_t}{(n-1)\textrm{$\sf{SNR}$}} \right) - F_G\left(  \frac{(2^{R}-1) L_t}{n \textrm{$\sf{SNR}$}} \right) \right). 
\end{equation}


\begin{figure}
\centering
\includegraphics[width=\textwidth]{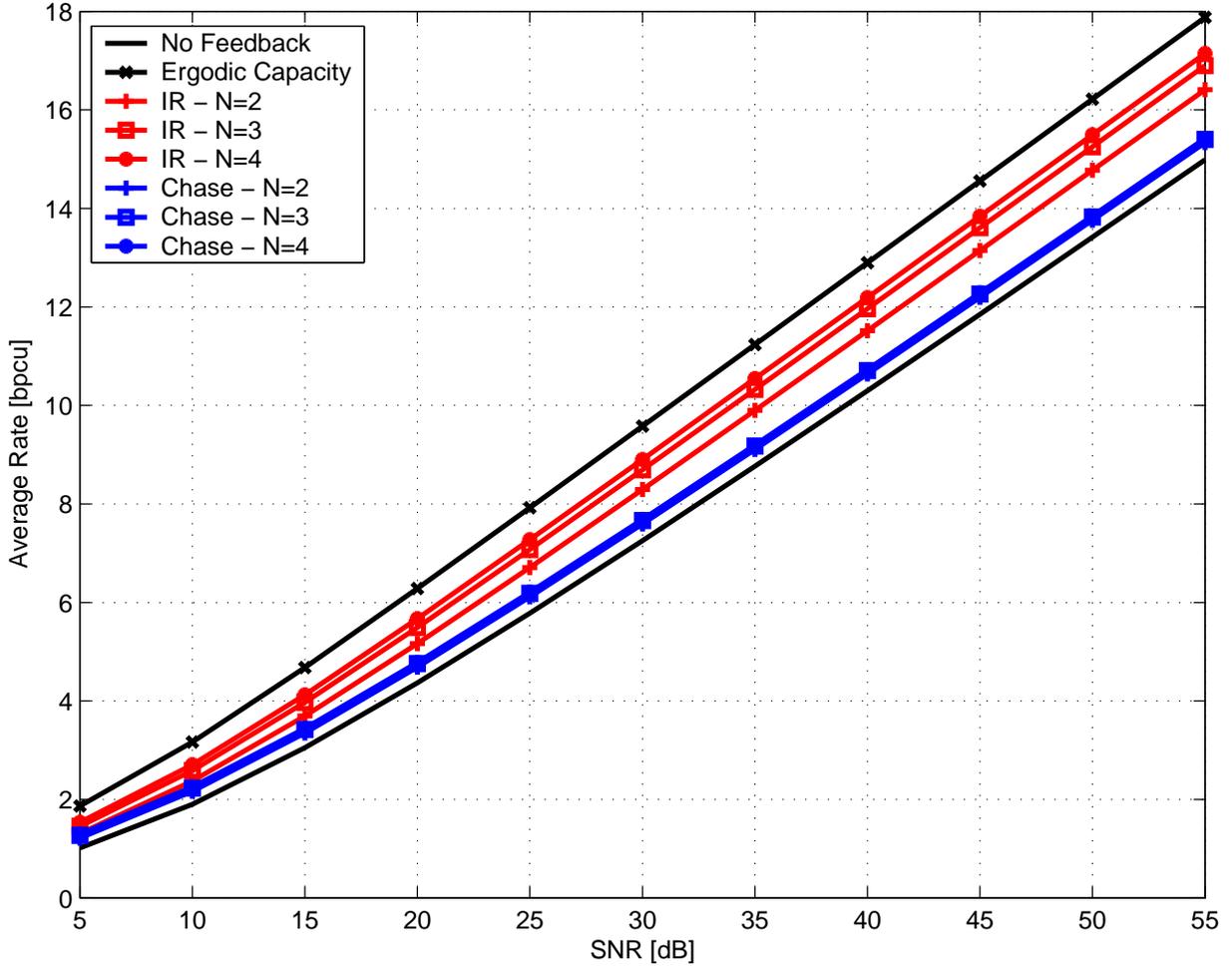}
\caption{The optimal average rate performance of IR and CC in a $(L_t=2, L_r=1)$ MISO Gaussian channel with different $N$.} \label{fig:MISO_Lt2}
\end{figure}

\begin{figure}
\centering
\includegraphics[width=\textwidth]{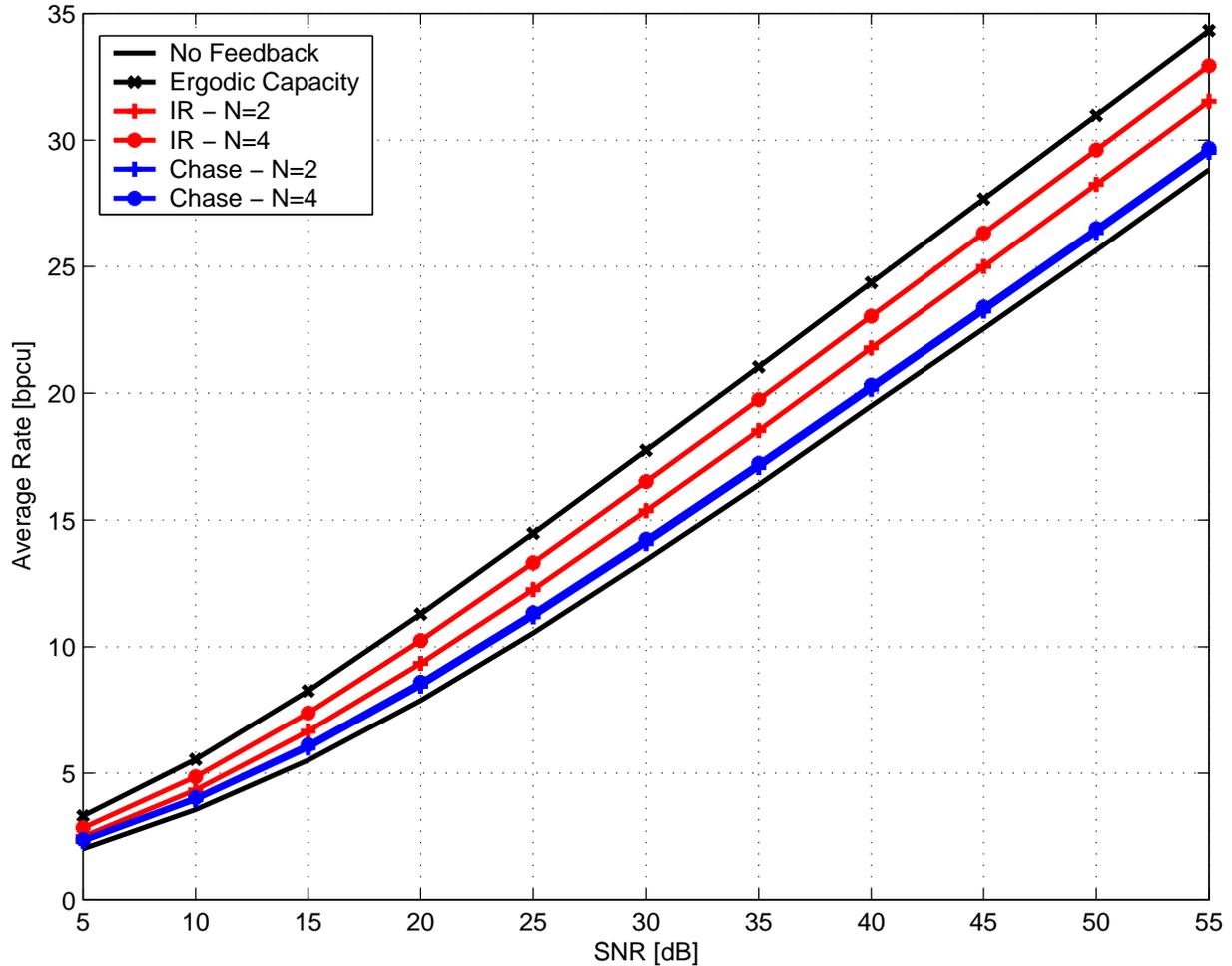}
\caption{The optimal average rate performance of IR and CC in a $(L_t=2, L_r=2)$ MIMO Gaussian channel.} \label{fig:MIMO_Lt2Lr2}
\end{figure}

Unfortunately, such analytically tractable expressions are
unavailable for the general MIMO channel, and we have resorted to
numerical methods to solve the optimization problems.
Fig.~\ref{fig:MISO_Lt2} and Fig.~\ref{fig:MIMO_Lt2Lr2} give the
numerical results of Problems (\ref{prb:IRavgrate_isotr})
(\ref{prb:CCavgrate_isotr}) for $(L_t=2, L_r=1)$ MISO and $(L_t=2,
L_r=2)$ MIMO systems, respectively. The ergodic capacity and the
optimal average rate of the no-feedback scheme are natural upper and
lower bounds of the ARQ schemes \cite{SLF:08}, respectively, and
they are plotted for reference. As one can see from the figures, IR
performs better than the CC and no-feedback scheme and is close to
the ergodic capacity. Moreover, the optimal average rate continues
to increase as more ARQ rounds are allowed. In fact, the average
rate of IR can be proved to asymptotically achieve the ergodic
capacity, regardless of the fading distribution \cite[Lemma
2]{SLF:08}. On the other hand, the performance of CC is very
limited, and increasing the maximum allowable ARQ rounds $N$ brings
almost negligible gain to the optimal average rate.

%

\section{Optimal LDC design: a mutual information analysis}
\label{sec:LDC_MI}

The previous section discusses the traditional IR and CC protocols
and evaluates their performance. The IR protocol has excellent
average rate thanks to two properties that allow full utilization of
the ARQ retransmission. The first is the flexibility in rate
assignment $\left\{ R_n \right\}_{n=1}^{N}$, which can be
accomplished by either the rate-compatible channel code via
puncturing
\cite{Hagenauer:88,HKM:04,SCV:04,VSW:05,HKKM:06,KHRM:06,HA:08} or
the recently developed rateless code
\cite{Luby:02,Shokrollahi:06,SVW:06}. On the other hand, the CC
protocol has only one rate parameter $R$ to optimize, and each
retransmission can reduce the rate by only an integer fraction.
However, one should note that the advantage of IR rate assignment
comes at the cost of complicated packetization: each transmission
could have very different packet length, which is undesirable in
practice. From this perspective, CC is preferred as each
retransmission is a simple repetition of the previous packet. The
second property is that each transmission in IR is
capacity-achieving, provided that the channel coding and space-time
transmission are carefully designed. For CC, however, it is clear
from (\ref{eqn:CCcap_gen}) that such capacity optimality is not
valid in general, due to the repetition nature of this protocol.
This also contributes to its poor performance.

This section introduces the MIMO-HARQ design based on STC, which
combines the advantages of both IR and CC. The STC-based protocol
has the IR capacity-achieving property for each retransmission,
while still enjoys the repetition nature of CC.
Information-theoretic tools will be used in this section for the STC
design of MIMO-HARQ. In Section~\ref{sec:LDC_error}, this design
problem will be revisited from the decoding error probability
perspective.

The idea can be best understood from the following simple example.
Consider $L_t=2$, $L_r=1$ and $N=2$. For such configuration, the
celebrated Alamouti code gives
\begin{equation}
\label{eqn:ala_exm}
\mathbf{X} = \left [ \begin{array}{cc}
s_1 & -s_2^* \\
s_2 & s_1^*
\end{array} \right ]
= \left[ \mathbf{x}_1, \mathbf{x}_2 \right].
\end{equation}
The traditional use of Alamouti code would transmit $\mathbf{X}$ in
two channel uses. With ARQ feedback, however, one can separately
transmit the first and second column. To be more specific,
$\mathbf{x}_1$ is sent first. The receiver gets $y_1 = h_1 s_1 + h_2
s_2 + z_1$ and jointly decodes $(s_1, s_2)$ in a maximum-likelihood
(ML) manner. In case the first decoding attempt fails, a NACK will
be sent back to the transmitter asking for the transmission of the
second column $\mathbf{x}_2$. With both columns transmitted, the
receiver can perform the usual Alamouti decoding to recover $s_1$
and $s_2$ in the second decoding attempt.

The idea behind this scheme is that if the channel matrix is
``nice'', it may be unnecessary to send a full-diversity ST
codeword\footnote{This is further discussed at the end of
Sec.~\ref{sec:LDC_error}.}. The first transmission, which only sends
$\mathbf{x}_1$, is an aggressive attempt to exploit the channel. If
the random channel is not good enough, we step back to the Alamouti
scheme, with the help of ARQ feedback. It should be mentioned that
HARQ schemes jointly considering STC and packet retransmission has
been studied in the literature, e.g., \cite{KMY:04,CP:08}, but none
of the designs are from an information-theoretic perspective and
hence do not give the best performance one can hope from STC-based
HARQ. On the other hand, the information-theoretic studies of such
schemes \cite{GCD:06,CGRC:08} are mostly focused on the high SNR
diversity-multiplexing-delay tradeoff and do not study the optimal
STC design for a finite SNR.



\subsection{General framework}
\label{sec:LDC_MI_fram}

The Linear Dispersion Code (LDC), proposed by Hassibi and Hochwald
\cite{HH:02}, is a general STC design that incorporates most of the
known STBCs as special cases. This tool is used and a LDC framework
is formulated to study the performance of STCs for the MIMO-HARQ
design. Consider a $(L_t, L_r)$ MIMO system. For each LDC codeword
$K$ modulated complex symbols are transmitted, and $T$ is used to
denote the maximum number of time intervals of the LDC codeword for
all ARQ rounds, i.e., $T=\sum_{n=1}^{N}T_n$. A LDC codeword can be
expressed as a linear combination of the symbols\footnote{As our
focus is to study the mutual information of the LDC structure with
HARQ, the effect of channel coding length $L_n$ is irrelevant and is
ignored in the discussion.}
\begin{eqnarray}
\mathbf{X} &=& \sum_{k=1}^{K} \alpha_k \mathbf{A}_k + j \beta_k \mathbf{B}_k \nonumber \\
&=& \sum_{k=1}^{K} s_k \mathbf{C}_k + s_k^* \mathbf{D}_k,
\end{eqnarray}
where $s_k = \alpha_k + j \beta_k$ are the modulated symbols, and $\mathbf{A}_k, \mathbf{B}_k, \mathbf{C}_k, \mathbf{D}_k \in \mathcal{C}^{L_t \times T}$ are the LDC spreading matrices with
\begin{eqnarray}
\mathbf{A}_k &=&  \mathbf{C}_k + \mathbf{D}_k, \nonumber \\
\mathbf{B}_k &=&  \mathbf{C}_k - \mathbf{D}_k.
\end{eqnarray}

For the HARQ transmission with a maximum $N$ rounds, the overall LDC codeword $\mathbf{X}$ shall be divided into $N$ sub-codewords
\begin{equation}
\mathbf{X} = \left[ \mathbf{X}_1, \mathbf{X}_2, \cdots, \mathbf{X}_{N} \right],
\end{equation}
where $\mathbf{X}_n \in \mathcal{C}^{L_t \times T_n}$ is for transmission at ARQ round $n$. The LDC spreading matrices can be similarly divided
\begin{eqnarray}
\mathbf{A}_k &=& \left[ \mathbf{A}_{k,1}, \mathbf{A}_{k,2}, \cdots, \mathbf{A}_{k,N} \right], \nonumber \\
\mathbf{B}_k &=& \left[ \mathbf{B}_{k,1}, \mathbf{B}_{k,2}, \cdots, \mathbf{B}_{k,N} \right],
\end{eqnarray}
or
\begin{eqnarray}
\mathbf{C}_k &=& \left[ \mathbf{C}_{k,1}, \mathbf{C}_{k,2}, \cdots, \mathbf{C}_{k,N} \right], \nonumber \\
\mathbf{C}_k &=& \left[ \mathbf{D}_{k,1}, \mathbf{D}_{k,2}, \cdots, \mathbf{D}_{k,N} \right],
\end{eqnarray}
where $\mathbf{A}_{k,n}, \mathbf{B}_{k,n}, \mathbf{C}_{k,n}, \mathbf{D}_{k,n} \in \mathcal{C}^{L_t \times T_n}$. Defining
\begin{eqnarray}
T^{(n)} &=& \sum_{i=1}^{n} T_i, \\
\mathbf{A}_k^{(n)} &=& \left[ \mathbf{A}_{k,1}, \mathbf{A}_{k,2}, \cdots, \mathbf{A}_{k,n} \right], \nonumber \\
\mathbf{B}_k^{(n)} &=& \left[ \mathbf{B}_{k,1}, \mathbf{B}_{k,2}, \cdots, \mathbf{B}_{k,n} \right], \nonumber \\
\mathbf{C}_k^{(n)} &=& \left[ \mathbf{C}_{k,1}, \mathbf{C}_{k,2}, \cdots, \mathbf{C}_{k,n} \right], \nonumber \\
\mathbf{D}_k^{(n)} &=& \left[ \mathbf{D}_{k,1}, \mathbf{D}_{k,2}, \cdots, \mathbf{D}_{k,n} \right],
\end{eqnarray}
the accumulated receive signal after the $n$-th transmission can be written as
\begin{equation}
\label{eqn:RXsig_n}
\mathbf{Y}^{(n)} = \sqrt{\frac{\textrm{$\sf{SNR}$}}{L_t}} \mathbf{H} \mathbf{X}^{(n)} +  \mathbf{Z}^{(n)},
\end{equation}
with
\begin{eqnarray}
\mathbf{X}^{(n)} &=& \left[ \mathbf{X}_1, \mathbf{X}_2, \cdots, \mathbf{X}_{n} \right] \nonumber  \\
&=& \sum_{k=1}^{K} \alpha_k \mathbf{A}_k^{(n)} + j \beta_k \mathbf{B}_k^{(n)} \label{eqn:LD_n} \\
&=& \sum_{k=1}^{K} s_k \mathbf{C}_k^{(n)} + s_k^* \mathbf{D}_k^{(n)}
\label{eqn:LD_n2}.
\end{eqnarray}

The capacity of the signal model (\ref{eqn:RXsig_n}) is denoted as
$C_{\textrm{$\sf{ld}$}}^{(n)}(\mathbf{H})$, which is determined by
the LD structure (\ref{eqn:LD_n}) or (\ref{eqn:LD_n2}). Meanwhile,
notice that the effective rate for ARQ round $n$ is
\begin{equation}
R^{(n)} = \frac{R}{T^{(n)}},
\end{equation}
where $R$ is the channel coding rate, which is a design parameter.
Now directly applying the general result (\ref{eqn:avgrate2}), the
average rate of LDC-based MIMO-HARQ can be expressed as
\begin{eqnarray}
\bar{R}_{\textrm{$\sf{ld}$}} &=& \sum_{n=1}^{N} \left( R^{(n)} - R^{(n+1)} \right)  \pr{ C_{\textrm{$\sf{ld}$}}^{(n)}\left( \mathbf{H} \right) \geq R^{(n)} } \nonumber  \\
&=&  \sum_{n=1}^{N} \frac{R T_{n+1}}{T^{(n)} T^{(n+1)}} \pr{ C_{\textrm{$\sf{ld}$}}^{(n)}\left( \mathbf{H} \right) \geq \frac{R}{T^{(n)}} }
\label{eqn:avgrateLDC}
\end{eqnarray}
where we use the convention $T^{(N+1)} \doteq \infty$ such that
$R^{(N+1)} = 0$, and $T_{N+1} / T^{(N+1)} = 1$.

The general problem of average rate maximization for LDC-based MIMO-HARQ can be formally casted as
\begin{equation}
\label{prb:LDavgrate_gen}
\begin{array}{ll}
\underset{R, \left\{  \mathbf{A}_n, \mathbf{B}_n, T_n \right\}_{n=1}^{N}}{\mbox{maximize}} &  \sum_{n=1}^{N} \frac{R T_{n+1}}{T^{(n)} T^{(n+1)}} \pr{ C_{\textrm{$\sf{ld}$}}^{(n)}\left( \mathbf{H} \right) \geq \frac{R}{T^{(n)}} } \\
\mbox{subject to } & R \geq 0 \\
{} & T_n \textrm{ is a positive integer} \\
{} & \sum_{n=1}^{N}T_n = T \\
{} & \trace{ \sum_{k=1}^{K}  \left(  \mathbf{A}_{k,n} \mathbf{A}_{k,n}^{H}  +  \mathbf{B}_{k,n} \mathbf{B}_{k,n}^{H} \right) } = 2 L_t T_n, \forall n=1, \cdots, N.
\end{array}
\end{equation}
Notice that $C_{\textrm{$\sf{ld}$}}^{(n)}\left( \mathbf{H} \right)$ implicitly relies on $ \left\{  \mathbf{A}_i, \mathbf{B}_i, T_i \right\}_{i=1}^{n}$. The power constraint in Problem (\ref{prb:LDavgrate_gen}) will be discussed in more detail in Section \ref{sec:LDC_MI_rem}.

\subsection{Design criterion}
\label{sec:LDC_MI_crit}

Directly solving Problem (\ref{prb:LDavgrate_gen}), either analytically or numerically, is difficult, but there are some key  properties that will help guide the optimal LDC design. The first observation is that the average rate is monotonic with the LDC capacity, and an achievable upper bound is the capacity of the MIMO channel. This leads to the following design criterion for the LDC design of MIMO-HARQ.

\begin{crit} [Capacity-based LDC design criterion]
\label{crit:cap_LDC}
The LDC should satisfy
\begin{equation}
C_{\textrm{$\sf{ld}$}}^{(n)}\left( \mathbf{H} \right) = C_{\textrm{$\sf{mimo}$}}\left( \mathbf{H}, \frac{\textrm{$\sf{SNR}$}}{L_t} \mathbf{I}_{L_t} \right)
\end{equation}
for any ARQ round $ n = 1, \cdots, N$.
\end{crit}

Notice that the intuition behind this criterion is that each LDC transmission should create an equivalent MIMO channel that is capacity lossless. Under the constraint of isotropic Gaussian distribution, this criterion gives the largest capacity for each transmission, which then minimizes the probability of decoding failure for any given $R$ and $ \left\{ T_n \right\}_{n=1}^{N}$.

With Criterion \ref{crit:cap_LDC}, the original Problem (\ref{prb:LDavgrate_gen}) is greatly simplified, as the dependence on $ \left\{  \mathbf{A}_n, \mathbf{B}_n \right\}_{n=1}^{N}$ disappears. The problem becomes how to find the optimal $R$ and $\left\{ T_n \right\}_{n=1}^{N}$ that maximizes the average rate. In most of the remaining work a special case of $T_n = 1$, $\forall n = 1, \cdots, N$ is considered. There are two reasons to focus on this special situation.
\begin{itemize}
\item [1)] Optimizing over all possible combinations of $\left\{ T_n \right\}_{n=1}^{N}$ is complex even for moderate $N$. This requires solving a single-variable optimization problem for each possible combination of $\left\{ T_n \right\}_{n=1}^{N}$ satisfying $\sum_{n=1}^{N}T_n = T$ and $T_n$ being a positive integer. The optimal solution may change with respect to the operating SNR or the channel distribution. In practice $\left\{ T_n \right\}_{n=1}^{N}$ may be pre-determined with the choice of LDC, which leaves only one variable $R$ to optimize over.
\item [2)] Choosing $T_n = 1$ ensures that each ARQ round is of minimum delay, which is an important advantage. Meanwhile, this minimum delay may also result in the best throughput, as it is empirically observed as a good balance between $\frac{ T_{n+1}}{T^{(n)} T^{(n+1)}}$ and $ \pr{ C_{\textrm{$\sf{ld}$}}^{(n)}\left( \mathbf{H} \right) \geq \frac{R}{T^{(n)}} }$. Intuitively, large $T_n$ will result in a marginal gain in the decoding successful probability, but a significant decrease of the multiplicative coefficient. 
\end{itemize}

To summarize, we give the following corollary.
\begin{coro}
\label{coro:cap_LDC}
Assume isotropic Gaussian input distribution over all available transmit antennas. For a given set of $\left\{ T_n \right\}_{n=1}^{N}$, the optimal average rate of a LDC-based MIMO-HARQ protocol is
\begin{equation}
\label{eqn:coro_cap_LDC1}
\bar{R}_{\textrm{$\sf{ld}$}}^{*} = \max_{R} \sum_{n=1}^{N} \frac{R T_{n+1}}{T^{(n)} T^{(n+1)}} \pr{ C_{\textrm{$\sf{mimo}$}}\left( \mathbf{H}, \frac{\textrm{$\sf{SNR}$}}{L_t} \mathbf{I}_{L_t} \right) \geq \frac{R}{T^{(n)}}}.
\end{equation}
In the special case of $T_n = 1$, $\forall n = 1, \cdots, N$, the optimal average rate becomes
\begin{equation}
\label{eqn:coro_cap_LDC2}
\bar{R}_{\textrm{$\sf{ld}$}}^{*} = \max_{R} \sum_{n=1}^{N-1} \frac{R}{n(n+1)} \pr{ C_{\textrm{$\sf{mimo}$}}\left( \mathbf{H}, \frac{\textrm{$\sf{SNR}$}}{L_t} \mathbf{I}_{L_t} \right) \geq \frac{R}{n}} +  \frac{R}{N} \pr{ C_{\textrm{$\sf{mimo}$}}\left( \mathbf{H}, \frac{\textrm{$\sf{SNR}$}}{L_t} \mathbf{I}_{L_t} \right) \geq \frac{R}{N}}.
\end{equation}
\end{coro}

Criterion \ref{crit:cap_LDC} and Corollary \ref{coro:cap_LDC} are based on the mutual information of the equivalent MIMO channel. To more conveniently evaluate the optimality of any given LDC structure, it is helpful to develop a criterion that explicitly relies on the LDC spreading matrices $\left\{ \mathbf{C}_k, \mathbf{D}_k  \right\}_{k=1}^{K}$, which lead to Theorem~\ref{thm:cap_LDC}.
\begin{theorem}
\label{thm:cap_LDC}
Assume $L_r \geq L_t$ and $K = L_t T$. Define
\begin{eqnarray}
\mathbf{U}^{(n)} &=& \left[ \vvec{\mathbf{C}_{1}^{(n)}}, \cdots, \vvec{\mathbf{C}_{K}^{(n)}} \right], \nonumber \\
\mathbf{V}^{(n)} &=& \left[ \vvec{\mathbf{D}_{1}^{(n)}}, \cdots, \vvec{\mathbf{D}_{K}^{(n)}} \right],
\end{eqnarray}
and
\begin{equation}
\mathbf{F}^{(n)} = \left[
\begin{array}{ll}
\mathbf{U}^{(n)} &\mathbf{V}^{(n)} \\
\conj{\mathbf{V}^{(n)}} & \conj{\mathbf{U}^{(n)}}
\end{array}
\right].
\end{equation}
The LDC-based HARQ protocol $\{ \mathbf{C}_{k,n}, \mathbf{D}_{k,n} \}_{n=1}^{N}{}_{k=1}^{K}$ leads to the optimal average rate performance if and only if
\begin{equation}
\mathbf{F}^{(n)} {\mathbf{F}^{(n)}}^{H} = \mathbf{I}_{2L_t T^{(n)}},
\end{equation}
for all $n=1, \cdots, N$.
\end{theorem}

The proof follows almost directly from \cite[Theorem 5.3.1]{ZLW:05}. The challenge is that Theorem \ref{thm:cap_LDC} requires to consider $K > L_t T^{(n)}$ for each $n=1, \cdots, N-1$, while \cite[Theorem 5.3.1]{ZLW:05} only holds for $K = L_t T$. However, a careful study of the proof of \cite[Theorem 5.3.1]{ZLW:05} shows that $\mathbf{F}^{(n)} {\mathbf{F}^{(n)}}^{H} = \mathbf{I}_{2L_t T^{(n)}}$ would be necessary and sufficient for the LDC to be average rate optimal.

Theorem \ref{thm:cap_LDC} considers the general LDCs that allow the
conjugation operation. This is a key feature in some codes, e.g.,
Orthogonal STBC, but it has been argued \cite{HP:02} that there is
not always a significant gain over complex LDCs, in which no
conjugation operation is allowed. Hence, it is instructive to give
the design criterion for complex LDCs, which follows directly from
Theorem \ref{thm:cap_LDC}.
\begin{coro}
\label{coro:cap_complexLDC}
For a complex LDC
\begin{equation}
\mathbf{X}^{(n)} = \sum_{k=1}^{K} s_k \mathbf{C}_k^{(n)}
\end{equation}
with $L_r \geq L_t$ and $K = L_t T$, it is average rate optimal if and only if
\begin{equation}
\mathbf{U}^{(n)} {\mathbf{U}^{(n)}}^{H} = \mathbf{I}_{L_t T^{(n)}},
\end{equation}
for all $n=1, \cdots, N$.
\end{coro}




\subsection{Evaluating existing LDCs}
\label{sec:LDC_MI_eval}

Previous sections analyzed the optimal average rate performance of
LDC-based MIMO-HARQ and proposed the design criterion of the
corresponding space-time structure. This section will attempt to use
this general framework to evaluate  existing LDC structures and
their usage in the MIMO-HARQ setting. Although Theorem
\ref{thm:cap_LDC} and Corollary \ref{coro:cap_complexLDC} provide
straightforward means to evaluate given LDCs, we shall proceed to
study the equivalent MIMO channel capacity and use Criterion
\ref{crit:cap_LDC} whenever it is feasible and simple. Directly
evaluating the equivalent MIMO channel capacity and comparing to the
physical channel capacity shed light on the (sub)optimality of the
LDC. We will start with the simple MISO $L_t=2, L_r=1, N=2$ setting,
in which analytical (sub)optimality can be rigorously shown. For the
more complicated MIMO settings, numerical simulations are performed
as a main tool to evaluate existing LDCs.


\subsubsection{MISO $L_t=2, L_r=1, N=2$}
This is the simplest MISO setting, in which the celebrated Alamouti code was invented \cite{Alamouti:98}. The following LDC-based HARQ protocols are studied. 

\textbf{Spatial Multiplexing with Repetition:}

This is the multiple-antenna version of the CC scheme. Spatial multiplexing vector $\mathbf{x} = [\mathbf{x}(1), \mathbf{x}(2)]^t$ is repeated upon a NACK, and the overall codeword after $N=2$ transmissions is
\begin{equation}\left [
\begin{array}{ll}
\mathbf{x}(1) & \mathbf{x}(1) \\
\mathbf{x}(2) & \mathbf{x}(2)
\end{array}
\right].
\end{equation}

Since ARQ round 1 is a spatial multiplexing, it is capacity-optimal:
\begin{eqnarray}
C_{\textrm{$\sf{sm}$}}^{(1)}\left( \mathbf{h}^t \right) &=& C_{\textrm{$\sf{mimo}$}}\left( \left[ \begin{array}{ll}
\mathbf{h}(1) & \mathbf{h}(2)
\end{array}\right], \frac{\textrm{$\sf{SNR}$}}{2} \mathbf{I}_{L_t} \right) \nonumber  \\
&=& \log \left( 1 + \frac{\textrm{$\sf{SNR}$}}{2} g_2 \right),
\end{eqnarray}
where $g_2$ is defined in (\ref{eqn:def_g_Lt}). The simple repetition in ARQ round 2, however, leads to a capacity loss:
\begin{eqnarray}
C_{\textrm{$\sf{sm}$}}^{(2)}\left( \mathbf{h}^t \right) &=& \frac{1}{2} C_{\textrm{$\sf{mimo}$}}\left( \left[ \begin{array}{ll}
\mathbf{h}(1) & \mathbf{h}(2) \\
\mathbf{h}(1) & \mathbf{h}(2)
\end{array}\right], \frac{\textrm{$\sf{SNR}$}}{2} \mathbf{I}_{L_t} \right)  \nonumber  \\
&=& \frac{1}{2} \log \left( 1 + \textrm{$\sf{SNR}$} g_2 \right) \nonumber  \\
&=& \frac{1}{2} C_{\textrm{$\sf{mimo}$}}\left( \mathbf{h}^t, \textrm{$\sf{SNR}$} \mathbf{I}_{L_t} \right) \nonumber \\
&<& C_{\textrm{$\sf{mimo}$}}\left( \mathbf{h}^t, \frac{\textrm{$\sf{SNR}$}}{2} \mathbf{I}_{L_t} \right),
\end{eqnarray}
for $\textrm{$\sf{SNR}$} > 0$, which proves the strict suboptimality of this protocol.

\textbf{Antenna Switching (AS):}

This scheme activates only one antenna at a time, and rotates the active antenna to achieve spatial diversity. The overall codeword is
\begin{equation}\left [
\begin{array}{ll}
x & 0 \\
0 & x
\end{array}
\right].
\end{equation}
This is apparently a low-rate suboptimal scheme with
\begin{eqnarray}
C_{\textrm{$\sf{as}$}}^{(1)}\left( \mathbf{h}^t \right) &=& \log \left( 1 + \textrm{$\sf{SNR}$} g_1 \right) < C_{\textrm{$\sf{mimo}$}}\left( \mathbf{h}^t, \frac{\textrm{$\sf{SNR}$}}{2} \mathbf{I}_{L_t} \right) \\
C_{\textrm{$\sf{as}$}}^{(2)}\left( \mathbf{h}^t \right) &=& \frac{1}{2} \log \left( 1 + \textrm{$\sf{SNR}$} g_2 \right) < C_{\textrm{$\sf{mimo}$}}\left( \mathbf{h}^t, \frac{\textrm{$\sf{SNR}$}}{2} \mathbf{I}_{L_t} \right)
\end{eqnarray}
and hence neither transmission is capacity-optimal.

\textbf{Alamouti Code:}

The overall codeword after  $N=2$ transmissions is
\begin{equation}\left [
\begin{array}{ll}
\mathbf{x}(1) & - \mathbf{x}(2)^* \\
\mathbf{x}(2) & \mathbf{x}(1)^*
\end{array}
\right].
\end{equation}
The first transmission is a spatial multiplexing, which is optimal. It is also well known that the overall Alamouti structure also achieves the channel capacity \cite{HH:02}. Hence,
\begin{equation}
C_{\textrm{$\sf{Alamouti}$}}^{(1)}\left( \mathbf{h}^t \right) = C_{\textrm{$\sf{Alamouti}$}}^{(2)}\left( \mathbf{h}^t \right) = C_{\textrm{$\sf{mimo}$}}\left( \mathbf{h}^t, \frac{\textrm{$\sf{SNR}$}}{2} \mathbf{I}_{L_t} \right),
\end{equation}
and this protocol satisfies Criterion \ref{crit:cap_LDC}.

\textbf{Cyclic Delay Diversity (CDD):}

CDD was proposed in an OFDM setting \cite{DK:01} but the code itself presents an interesting space-time structure. It cyclically rotates the previously transmit column such that each symbol is sent from different antennas in different times slots. The overall codeword in the $\left( L_t=2, L_r=1 \right)$ MISO is
\begin{equation}\left [
\begin{array}{ll}
\mathbf{x}(1) & \mathbf{x}(2) \\
\mathbf{x}(2) & \mathbf{x}(1)
\end{array}
\right].
\end{equation}
ARQ round 1 again is a spatial multiplexing, and hence is optimal. For ARQ round 2, the equivalent channel matrix is
\begin{equation}
\mathbf{H}_{\textrm{$\sf{eq, cdd}$}} = \left [
\begin{array}{ll}
\mathbf{h}(1) & \mathbf{h}(2) \\
\mathbf{h}(2) & \mathbf{h}(1)
\end{array}
\right],
\end{equation}
and
\begin{eqnarray}
C_{\textrm{$\sf{cdd}$}}^{(2)}\left( \mathbf{h}^t \right) &=& \frac{1}{2} \log \det \left( \mathbf{I}_2 +  \frac{\textrm{$\sf{SNR}$}}{2} \mathbf{H}_{\textrm{$\sf{eq, cdd}$}} \mathbf{H}_{\textrm{$\sf{eq, cdd}$}}^{H} \right) \nonumber  \\
&=& \frac{1}{2} \log \det \left[
\begin{array}{cc}
1+ \frac{\textrm{$\sf{SNR}$}}{2} g_2 & \textrm{$\sf{SNR}$} \ \re{ \mathbf{h}(1)\mathbf{h}(2)^* } \\
\textrm{$\sf{SNR}$} \ \re{ \mathbf{h}(1)\mathbf{h}(2)^* } & 1+
\frac{\textrm{$\sf{SNR}$}}{2} g_2
\end{array}
\right] \nonumber  \\
&=& \frac{1}{2} \log\left( 1+\frac{\textrm{$\sf{SNR}$}}{2} | \mathbf{h}(1) + \mathbf{h}(2) |^2 \right) + \frac{1}{2} \log\left( 1+\frac{\textrm{$\sf{SNR}$}}{2} | \mathbf{h}(1) - \mathbf{h}(2) |^2 \right) \nonumber  \\
&\leq& \log\left( 1+\frac{\textrm{$\sf{SNR}$}}{2} \left( | \mathbf{h}(1) |^2 + | \mathbf{h}(2) |^2 \right) \right) \nonumber  \\
&=& C_{\textrm{$\sf{mimo}$}}\left( \mathbf{h}^t, \frac{\textrm{$\sf{SNR}$}}{2} \mathbf{I}_{L_t} \right).
\end{eqnarray}
This suggests that the CDD-based protocol is suboptimal. 

\textbf{Numerical Simulations:}

\begin{figure}
\centering
\includegraphics[width=\textwidth]{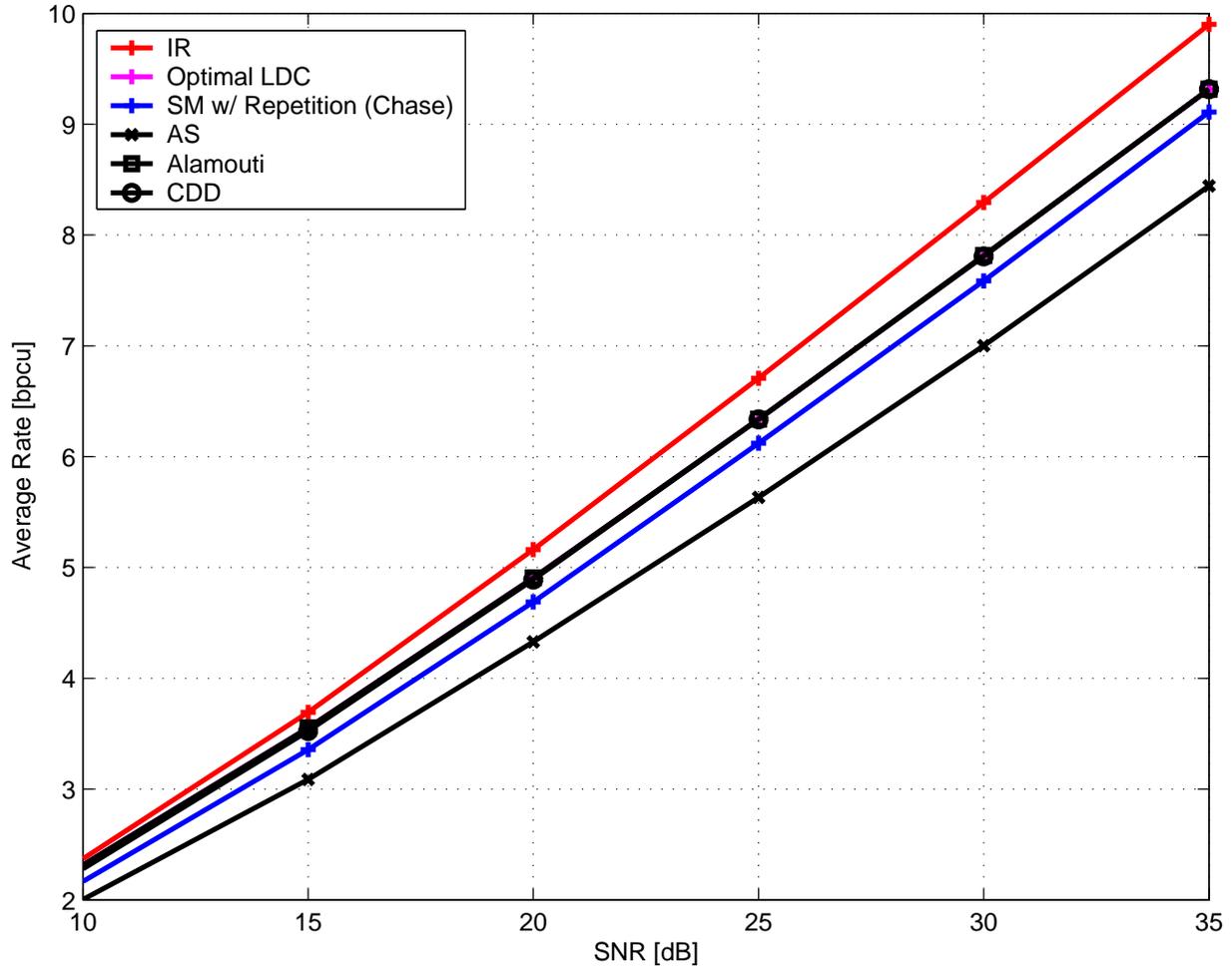}
\caption{The optimal average rate performance of several existing LDCs in a $(L_t=2, L_r=1, N=2)$ Gaussian MISO-HARQ channel.} \label{fig:Plot4_MISO_Lt2N2}
\end{figure}

In addition to the previous analysis, numerical optimization is
performed to obtain the average rates of several LDCs and the
results are reported in Fig.~\ref{fig:Plot4_MISO_Lt2N2}. Optimal
LDC, which satisfies Criterion \ref{crit:cap_LDC}, and IR are
plotted as references. It is interesting to observe that although
both AS and CDD are analytically proved to be sub-optimal, the
average rate performance of CDD is very close to the optimum (almost
negligible performance loss), while AS is extremely sub-optimal.
This is no surprise as CDD utilizes the spatial degrees of freedom
in a more efficient way than AS. SM-based repetition gives the
performance of CC and is not far away from the optimal LDC-based
protocol. Alamouti, as predicted in the analysis, is  the optimal
LDC.

\subsubsection{Full MIMO}

Analytically proving (sub)optimality becomes infeasible in most of
the full MIMO settings, as the comparison of mutual information of a
given LDC $\left\{  \mathbf{A}_n, \mathbf{B}_n \right\}_{n=1}^{N}$
to the MIMO channel capacity is difficult to make even for moderate
$N$. This section thus focuses on numerical simulations to get
insight on existing LDCs.

\begin{figure}
\centering
\includegraphics[width=\textwidth]{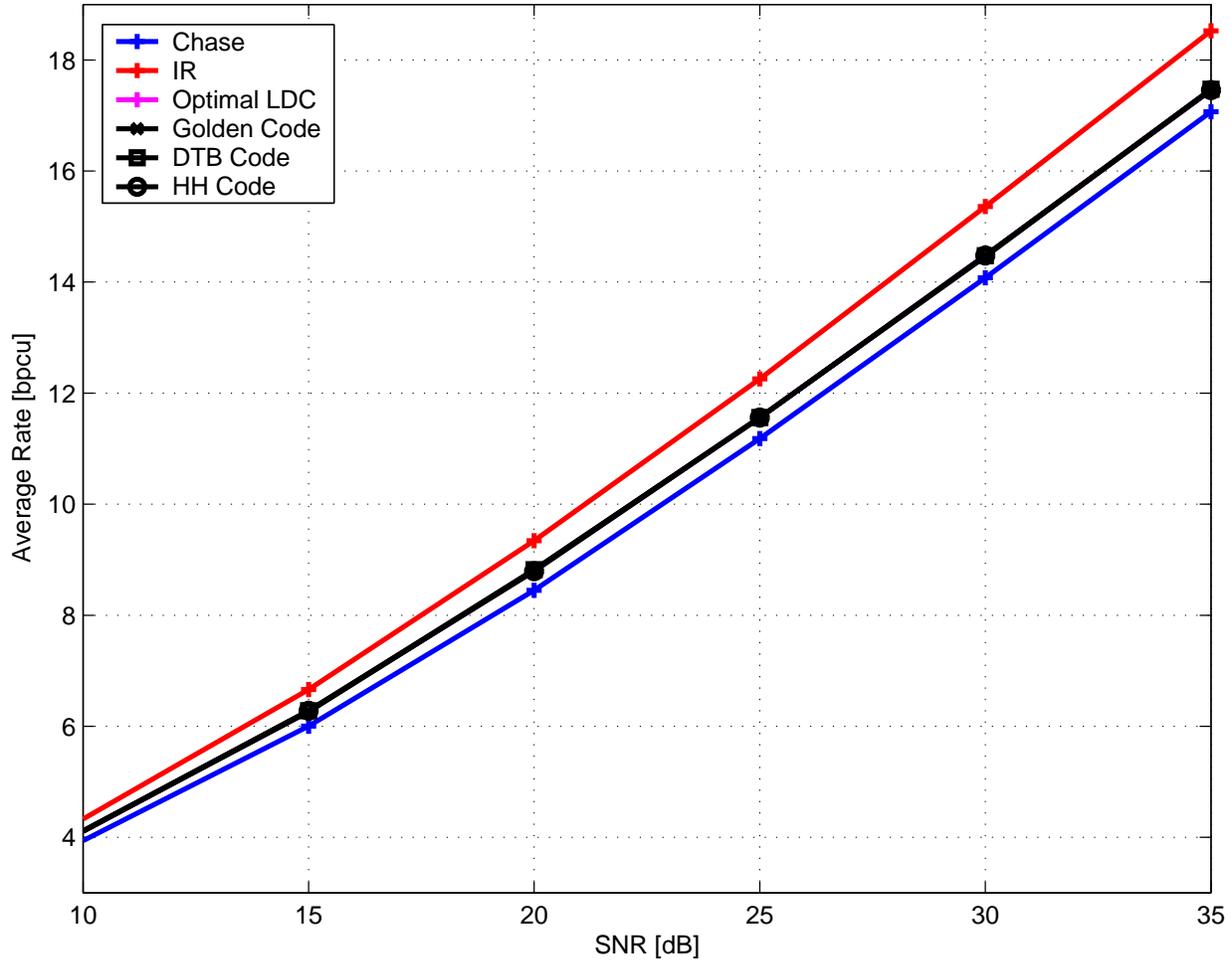}
\caption{The optimal average rate performance of several existing LDCs in a $(L_t=2, L_r=2, N=2)$ Gaussian MIMO-ARQ channel.} \label{fig:Plot5_MIMO_Lt2Lr2N2}
\end{figure}

Fig.~\ref{fig:Plot5_MIMO_Lt2Lr2N2} plots three well-known LDCs in a
$\left( L_t=2, L_r=2 \right)$ MIMO channel with $N=2$: the Golden
Code (Golden) \cite{BRV:05}, Damen, Tewfik and Belfiore's code (DTB)
\cite{DTB:02}, and Hassibi and Hochwald's LD code (HH)
\cite[Equation (31)]{HH:02}. These three LDCs are known in the
literature to be capacity optimal, and in fact it can be
proven\footnote{This can be done by direcly verifying Corollary
\ref{coro:cap_complexLDC} or Theorem \ref{thm:cap_LDC} using their
corresponding LDC spreading matrices.} that they satisfy Corollary
\ref{coro:cap_complexLDC} or Theorem \ref{thm:cap_LDC}, and is
average rate optimal as well (the first column of the codeword is
capacity lossless). Similar to the mutual information maximization
approach of LDC design \cite{HH:02}, typically there exist many LDCs
that  satisfy Criterion 1. In practice, one can search among these
codes to further consider other criteria \cite{HP:02}, such as the
diversity and coding gain.

\begin{figure}
\centering
\includegraphics[width=\textwidth]{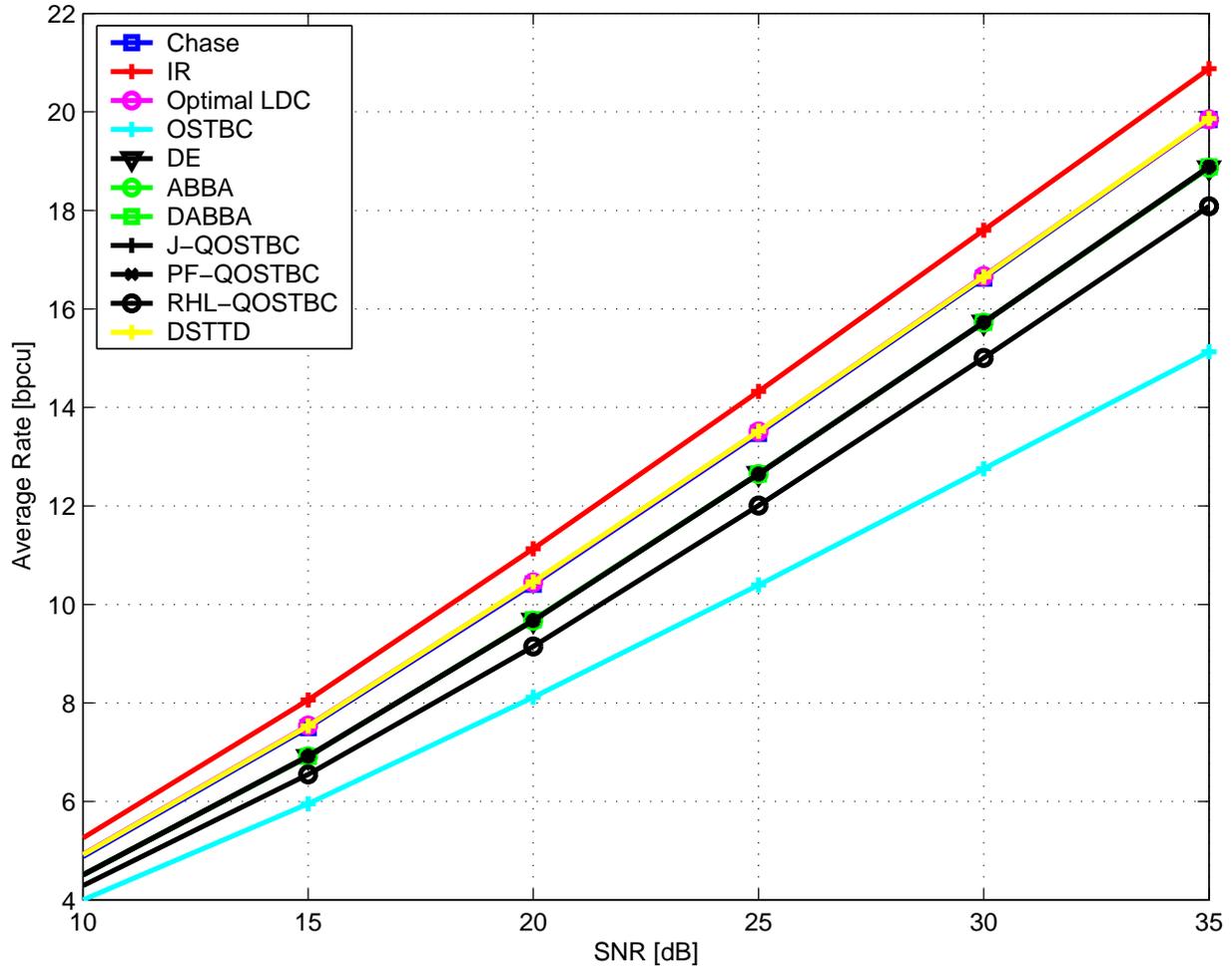}
\caption{The optimal average rate performance of several existing LDCs in a $(L_t=4, L_r=2, N=2)$ Gaussian MIMO-ARQ channel.} \label{fig:Plot6_MIMO_Lt4Lr2N2}
\end{figure}

\begin{figure}
\centering
\includegraphics[width=\textwidth]{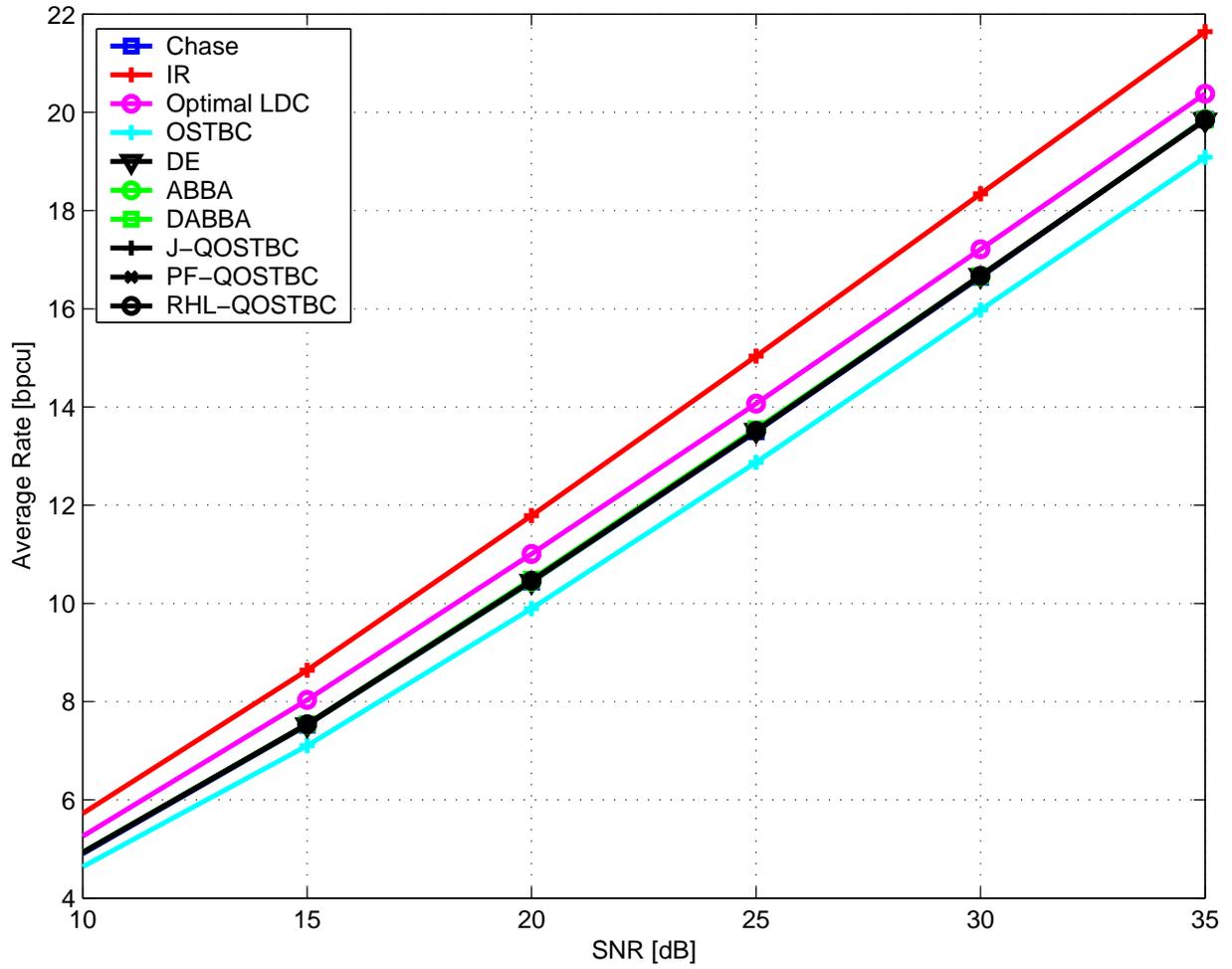}
\caption{The optimal average rate performance of several existing LDCs in a $(L_t=4, L_r=2, N=4)$ Gaussian MIMO-ARQ channel.} \label{fig:Plot7_MIMO_Lt4Lr2N4}
\end{figure}

For a $\left(L_t=4, L_r=2\right)$ MIMO-HARQ system,
Figs.~\ref{fig:Plot6_MIMO_Lt4Lr2N2} and
\ref{fig:Plot7_MIMO_Lt4Lr2N4} give the simulation results for $N=2$
and $N=4$, respectively. The LDCs under consideration are:
Orthogonal STBC (OSTBC) \cite[Chapter 7.6]{CN:06}, Diveristy
Embedded STC (DE) \cite{DCDA:08}, ABBA code \cite{TBH:00}, Double
ABBA code (DABBA) \cite{HT:02}, Jafarkhani's quasi-orthogonal STBC
(J-QOSTBC) \cite{Jafarkhani:01}, Papadias and Foschini's QOSTBC
(PF-QOSTBC) \cite{PF:03}, Ran, Hou and Lee's QOSTBC (RHL-QOSTBC)
\cite{RHL:03}, and Double Space-Time Transmit Diversity (DSTTD)
\cite{TI:01}. CC, IR and optimal LDC are again plotted as reference
curves.

Most existing LDCs for $L_t=4$ are obtained using $L_t=2$ LDCs as
building blocks, which use $2$ time slots. Hence, unlike other
examples in this paper, the HARQ protocols for $N=2$ use $T_1=T_2=2$
and the results are reported in Fig.~\ref{fig:Plot6_MIMO_Lt4Lr2N2}.
The first observation is that even the optimal LDC has an average
rate that is very close to CC. This suggests that with this specific
configuration, one should not expect much average rate gain from LDC
over simple repetition protocol. However, a well designed LDC-based
protocol may have other advantages, such as diversity gain and
simple decoding. The second conclusion is that among all the LDCs,
only DSTTD performs close to the optimal LDC\footnote{It should be
noted that  DSTTD is in fact capacity suboptimal.}. Among the other
LDCs, five (ABBA, DABBA, DE, J-QOSTBC and PF-QOSTBC) have very
similar optimal average rate, while RHL-QOSTBC is worse and OSTBC
gives the worst performance. The extreme suboptimality of OSTBC is
again due to its low-rate property (rate 3/4).

The conclusions are different in the case of $N=4$, in which each
ARQ round sends one column of the overall ST codeword matrix. From
Fig.~\ref{fig:Plot7_MIMO_Lt4Lr2N4}, there is a notable gain of
optimal LDC over CC. Meanwhile, none of the above
LDCs\footnote{DSTTD only uses two time slots and thus cannot be used
with $N=4$.} are optimal, which calls for the search of novel LDCs
that approach the performance of optimal LDC.

\subsection{Remarks}
\label{sec:LDC_MI_rem}

\begin{remark}
Power Constraint
\end{remark}

We would like to constrain the long-term average transmit power in
Problem (\ref{prb:LDavgrate_gen}). Two different types of power
constraints can be considered.
\begin{itemize}
\item [1)] \textbf{Each ARQ round has the same power.}

In this constraint, regardless of how many transmissions actually
take place, the overall transmit power is constant due to the equal
power allocation over ARQ rounds. The power constraint in the
original problem (\ref{prb:LDavgrate_gen}) falls into this category:
\begin{equation}
\label{eqn:PC1_1}
\trace{ \sum_{k=1}^{K}  \left(  \mathbf{A}_{k,n} \mathbf{A}_{k,n}^{H}  +  \mathbf{B}_{k,n} \mathbf{B}_{k,n}^{H} \right) } = 2 L_t T_n, \forall n=1, \cdots, N.
\end{equation}
Note that with this constraint, each transmit antenna or each time
slot can have different transmit power. Also the real and imaginary
parts of the transmit constellation can have unequal power.

Similar to \cite{HH:02}, more stringent power constraint can be posed to replace (\ref{eqn:PC1_1}):
\begin{equation}
\label{eqn:PC1_2}
\trace{\mathbf{A}_{k,n} \mathbf{A}_{k,n}^{H}} = \trace{\mathbf{B}_{k,n} \mathbf{B}_{k,n}^{H}} = \frac{L_t T_n}{K},
\end{equation}
or
\begin{equation}
\label{eqn:PC1_3}
\mathbf{A}_{k,n} \mathbf{A}_{k,n}^{H} = \mathbf{B}_{k,n} \mathbf{B}_{k,n}^{H} = \frac{T_n}{K} \mathbf{I}_{L_t},
\end{equation}
where (\ref{eqn:PC1_2}) ensures real and imaginary parts are transmitted with the same power, while (\ref{eqn:PC1_3}) further forces energy to be spread equally in all spatial and temporal dimensions. These power constraints can be incorporated into Problem (\ref{prb:LDavgrate_gen}).

\item [2)] \textbf{Dynamic power allocation among ARQ rounds.}

Instead of always having a constant power for each ARQ round, one can further allow power allocation among the $N$ ARQ rounds. For example, giving more power to ARQ round 1 will increase the probability that the first transmission succeeds. However this comes at the price of decreasing the power of potential ARQ rounds 2 to $N$, if the overall average transmit power is kept constant. There is a serious challenge to maintain a constant average power. The reason is that except for the first transmission, ARQ rounds 2 to $N$ happen only with a non-one probability. This challenge has been addressed in the scalar channel case \cite{SLF:08}, and the solution is to average the power consumption with respect to the channel fading distribution.

Assuming that ARQ round $n$ has a total transmit power $\rho_n$, the actually consumed power $\rho$ is a random variable with PMF
\begin{equation}
\label{eqn:PC2_1}
\rho = \left\{ \begin{array}{ll}
\frac{\sum_{i=1}^{n} \rho_i T_i}{\sum_{i=1}^{n} T_i}, & \textrm{if $\overline{\mathcal{A}_1}, \cdots, \overline{\mathcal{A}_{n-1}}, \mathcal{A}_n$; $\forall n=1, \cdots, N-1$}\\
\frac{\sum_{i=1}^{N} \rho_i T_i}{\sum_{i=1}^{N} T_i}, & \textrm{if $\overline{\mathcal{A}_1}, \cdots, \overline{\mathcal{A}_{N-1}}$}.
\end{array} \right.
\end{equation}
Note that dynamic power allocation also affects the equivalent MIMO channel capacity after ARQ round $n$: $C_{\textrm{$\sf{ld}$}}^{(n)}(\mathbf{H})$ is a TDMA-type one where different portions of the code have different power \cite{SLF:08}.

To maintain an average total transmit power $\textrm{$\sf{SNR}$}$, $\left(\rho_1, \cdots, \rho_{N} \right)$ should satisfy
\begin{equation}
\label{eqn:PC2_2}
\sum_{n=1}^{N-1} \frac{\sum_{i=1}^{n} \rho_i
T_i}{\sum_{i=1}^{n} T_i} \pr{ \overline{\mathcal{A}_1}, \cdots,
\overline{\mathcal{A}_{n-1}}, \mathcal{A}_n } + \frac{\sum_{i=1}^{N}
\rho_i T_i}{\sum_{i=1}^{N} T_i} \pr{ \overline{\mathcal{A}_1},
\cdots, \overline{\mathcal{A}_{N-1}} } = 1.
\end{equation}
\end{itemize}

\begin{remark}
Chase, LDC, and IR
\end{remark}

Let us revisit the average rate  for IR (\ref{prb:IRavgrate_isotr}), CC (\ref{prb:CCavgrate_isotr}), and LDC (\ref{prb:LDavgrate_gen}). It is straightforward to verify that if the rate assignment in IR is
\begin{equation}
\label{eqn:IR_LDC_rate_comp}
R^{(n)} = \frac{R}{T^{(n)}}, \quad T^{(n)} \textrm{ is a positive integer}, \quad  \forall n=1, \cdots, N,
\end{equation}
the average rate expression of IR and LDC will be the same, provided that Criterion \ref{crit:cap_LDC} is satisfied. This demonstrates that the advantages of IR over LDC is the flexibility in rate assignment for each ARQ round, which comes at the cost of unequal packetization.

In reality, nevertheless, the freedom in rate assignment of IR is very limited due to practical limitations. Considerations such as packetization or \emph{slotted} multi-user transmission usually ask for constant-length retransmissions, which leads to the same IR rate assignment as in LDC. With this constraint, the optimal average rates of IR and LDC are the same. In such cases, LDC has the potential advantage of receiver complexity over IR. Recall that IR requires \emph{code combining} at the receiver -- punctured codeword symbols need to be combined with the previous transmissions for another decoding. LDC-based protocol, on the other hand, only needs \emph{packet-level combining}. Thus, the receiver complexity is shifted from channel decoding to front-end demodulation/detection. If the LDC is well designed, such detection complexity can be moderate, e.g., the Alamouti code. This can be an important advantage of LDC.

At the same time, LDC can also be viewed as an enhanced CC scheme. The reason is that in the LDC-based protocol each retransmission is essentially just a repetition of previously transmitted symbols -- no new information is sent. However, the ``repetition'' in LDC is handled in a smart way such that the equivalent channel capacity is increased to approach the MIMO channel capacity, while the simple repetition CC scheme suffers from capacity loss. This can also be explained from the diversity perspective, which will be further developed in Sec.~\ref{sec:LDC_error}. The CC protocol naively repeats the same codeword upon each NACK, which cannot efficiently exploit the spatial diversity. A good LDC-based HARQ protocol spreads out the symbols onto different antennas upon each retransmission request to better exploit the spatial diversity. This becomes especially important in a quasi-static fading channel, as there is no time diversity to exploit.

\begin{remark}
Switching Between Multiplexing and Diversity
\end{remark}

Compared to the STC-based HARQ scheme, there is a similar work \cite{HP:05} on switching between multiplexing and diversity, which is worth some comments. It is  easy to see that the Alamouti example essentially performs a switching between multiplexing and diversity, just as \cite{HP:05} does. However, one can see that the ARQ-feedback-based approach is superior to the method in \cite{HP:05}. With ARQ feedback one actually sets the \emph{default} choice to be full multiplexing. Hence, data communication takes place without even knowing whether one should use multiplexing or diversity. Switching to the diversity scheme only happens whenever necessary. This is possible because full multiplexing is embedded in any diversity schemes, which is not exploited in \cite{HP:05}. For the scheme in \cite{HP:05}, the transmitter needs to wait for the feedback to inform which scheme to use \emph{before} the data communication can take place. Another advantage of our approach is that ARQ feedback informs the \emph{decoding status} to the transmitter, and the decision of switching to diversity scheme is made only when \emph{the receiver fails to decode the full multiplexing transmission}. This is better than the selection criteria of \cite{HP:05}, which use the minimum Euclidean distances and Demmel condition number of the MIMO channel to make the decision. Notice that the ultimate goal is to improve the decoding error performance, and hence decoding failure should be the best decision metric to determine switching. Our approach directly relies on the decoding error event, while \cite{HP:05} uses some indirect performance measures, which can only approximate the decoding error event.

\section{Optimal LDC design: an error probability analysis}
\label{sec:LDC_error}

Space-time codes design has followed two related but different paths. The original approach is from the viewpoint of detection theory. The focus is on studying the pariwise error probability (PWEP) \cite{Gea:96,Gea:99,TSC:98,SFG:02}, which has led to the well-known \emph{rank} and \emph{determinant} criteria that use \emph{diversity} and \emph{coding} gain to compare different STCs. A different approach is from the standpoint of information theory. It views STC as a modulation and study the optimal structure that preserves the channel mutual information \cite{Massey:02,HH:02}. There are also works that combine these two views \cite{HP:05,ZLW:05}. The previous section investigated the optimal LDC design from an information-theoretic point of view. In this section, an error probability analysis of the optimal LDC-based HARQ protocol is performed, and the corresponding design criterion is presented.

\subsection{$n$-th pairwise error probability}
\label{sec:LDC_error_fram}

Assume that the signal vector $\mathbf{s} = \left[ s_1, \cdots, s_K \right]^{t}$ is chosen from a uniformly distributed set $\mathcal{S}$ with cardinality $M$:
\begin{equation}
\mathcal{S} = \left\{ \mathbf{s}_0, \cdots, \mathbf{s}_{M-1} \right\}.
\end{equation}
The probability of a decoding error after the $n$-th ARQ round can be written as
\begin{eqnarray}
P_{e}^{(n)} &=& \frac{1}{M} \sum_{j=0}^{M-1} \pr{ \mathbf{s}^{(1)} \neq \mathbf{s}_{j}, \cdots, \mathbf{s}^{(n)} \neq \mathbf{s}_{j} | \mathbf{s}_{j} \textrm{ was sent}  } \label{eqn:P_e_n_1} \\
&=& \frac{1}{M} \mathbb{E}\left[ \sum_{j=0}^{M-1} \pr{ \bigcup_{\substack{i_1=0 \\i_1 \neq j}}^{M-1} \cdots \bigcup_{\substack{i_n=0 \\i_n \neq j}}^{M-1} \left\{ Q_{i_1, j}^{(1)}<0, \cdots, Q_{i_n, j}^{(n)}<0 \right\} | \mathbf{H}, \mathbf{s}_{j} } \right] \label{eqn:P_e_n_2} \\
& \leq & \frac{1}{M} \sum_{j=0}^{M-1} \sum_{\substack{i_1=0 \\i_1 \neq j}}^{M-1} \cdots \sum_{\substack{i_n=0 \\i_n \neq j}}^{M-1} \mathbb{E}\left[ \pr{ Q_{i_1, j}^{(1)}<0, \cdots, Q_{i_n, j}^{(n)}<0  | \mathbf{H}, \mathbf{s}_{j} } \right] \label{eqn:P_e_n_3}
\end{eqnarray}
where (\ref{eqn:P_e_n_3}) comes from the traditional union bound \cite{Fitz:07}, $\mathbf{s}_{j}$ is the transmitted vector, $\mathbf{s}^{(n)}$ is the detected vector after ARQ round $n$, $Q_{i,j}^{(n)}$ is the pairwise decision metric in the $n$-th ARQ round, which is defined in the following. The optimum ML decoding rule for deciding between two possible codewords $\mathbf{X}^{(n)}(\mathbf{s}_{i})$ and $\mathbf{X}^{(n)}(\mathbf{s}_{j})$ for a given channel realization $\mathbf{H}$ and the receiver observation $\mathbf{Y}^{(n)}$ is
\begin{equation}
|| \mathbf{Y}^{(n)} -  \sqrt{\frac{\textrm{$\sf{SNR}$}}{L_t}}\mathbf{H} \mathbf{X}^{(n)}(\mathbf{s}_{i}) ||_F^{2}
\begin{array}{l}
\stackrel{\mathbf{s}_{i}}{<} \\
\stackrel{\textstyle >}{\scriptstyle \mathbf{s}_{j}}
\end{array}
|| \mathbf{Y}^{(n)} -  \sqrt{\frac{\textrm{$\sf{SNR}$}}{L_t}}\mathbf{H} \mathbf{X}^{(n)}(\mathbf{s}_{j}) ||_F^{2}.
\end{equation}
Define the ML metric
\begin{equation}
V_{i}^{(n)} = || \mathbf{Y}^{(n)} -  \sqrt{\frac{\textrm{$\sf{SNR}$}}{L_t}}\mathbf{H} \mathbf{X}^{(n)}(\mathbf{s}_{i}) ||_F^{2},
\end{equation}
and $Q_{i,j}^{(n)}$ is defined as
\begin{equation}
Q_{i,j}^{(n)} \doteq V_{i}^{(n)} - V_{j}^{(n)}.
\end{equation}

For $n=1$, (\ref{eqn:P_e_n_3}) reduces to the traditional union bound for STC:
\begin{equation}
P_e \leq \frac{1}{M} \sum_{j=0}^{M-1} \sum_{\substack{i=0 \\i \neq j}}^{M-1}\mathbb{E}\left[ \pr{ Q_{i,j} < 0 | \mathbf{H}, \mathbf{s}_{j} } \right].
\end{equation}
For Rayleigh fading,
\begin{equation}
\pr{ Q_{i,j} < 0 | \mathbf{H}, \mathbf{s}_{j} } = Q\left( \sqrt{\frac{d_E^2(i,j)}{2}} \right)
\end{equation}
where $d_E^2(i,j)$ is the squared Euclidean distance between two received codewords $ \mathbf{Y}^{(n)}_{i}$ and $\mathbf{Y}^{(n)}_{j}$, with $\mathbf{Y}^{(n)}_{i} \doteq  \sqrt{\frac{\textrm{$\sf{SNR}$}}{L_t}} \mathbf{H} \mathbf{X}^{(n)}(\mathbf{s}_{i})$. Tools such as the Chernoff bound \cite{Gea:99,TSC:98} or the tighter bound in \cite{SFG:02} can be applied. Meanwhile, assuming high SNR would further simplify the upper bound and it has led to the rank and determinant criteria.

\begin{figure}
\centering
\includegraphics[width=0.2\textwidth]{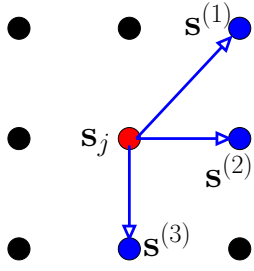}
\caption{An example illustrating different error events in different ARQ rounds. $\mathbf{s}_{j}$ (red) is transmitted, and $\mathbf{s}^{(n)}$ is the ML detector output at ARQ round $n$, which are all erroneous (blue).} \label{fig:Plot8_nPWEP}
\end{figure}

The performance analysis of MIMO-HARQ, however, is much more involved. The first complication comes from (\ref{eqn:P_e_n_3}). Unlike STC where only the PWEP is needed, analyzing MIMO-HARQ requires the \emph{$n$-th pairwise error probability ($n$-PWEP)}:
\begin{equation}
\label{eqn:nPWEP}
\pr{ Q_{i_1, j}^{(1)}<0, \cdots, Q_{i_n, j}^{(n)}<0  | \mathbf{H}, \mathbf{s}_{j} },
\end{equation}
in which the decision metrics $\left\{ Q_{i_l, j}^{(l)} \right\}_{l=1}^{n}$ are \emph{correlated} random variables. The complication is also intuitively understandable. The error event under consideration is when all the $n$ decoding attempts are erroneous. However, the decoding errors in different ARQ rounds could be very different (especially for short codes), depending on the channel realization and instantaneous noise.  Fig.~\ref{fig:Plot8_nPWEP} is one such example.

To analyze the $n$-PWEP, let us start with the statistics of $Q_{i,j}^{(n)}$. To simplify the derivation, define
\begin{equation}
\mathbf{D}^{(n)}_{i,j} \doteq  \sqrt{\frac{\textrm{$\sf{SNR}$}}{L_t}} \mathbf{H} \left( \mathbf{X}^{(n)}(\mathbf{s}_{i}) - \mathbf{X}^{(n)}(\mathbf{s}_{j}) \right),
\end{equation}
and
\begin{eqnarray}
\mathbf{y}^{(n)}_{i} &\doteq& \vvec{\mathbf{Y}^{(n)}_{i}}, \nonumber \\
\mathbf{d}^{(n)}_{i,j} &\doteq& \vvec{\mathbf{D}^{(n)}_{i,j}} = \mathbf{y}^{(n)}_{i} - \mathbf{y}^{(n)}_{j}, \nonumber \\
\mathbf{z}^{(n)} &\doteq& \vvec{\mathbf{Z}^{(n)}}.
\end{eqnarray}
Now conditioning on $\mathbf{H}$ and assuming $\mathbf{s}_{j}$ is transmitted, we have
\begin{eqnarray}
Q_{i,j}^{(n)} &=& || \mathbf{D}^{(n)}_{i,j}  + \mathbf{Z}^{(n)} ||_F^2 - ||\mathbf{Z}^{(n)} ||_F^2 \nonumber \\
&=& || \mathbf{D}^{(n)}_{i,j} ||_F^2 + 2 \re{ \trace{ \mathbf{D}^{(n)}_{i,j} { \mathbf{Z}^{(n)}}^H } } \label{eqn:Qij_n} \\
& = & {d_E^{(n)}(i,j)}^2 + W^{(n)}_{i,j}
\end{eqnarray}
where ${d_E^{(n)}(i,j)}^2 = || \mathbf{D}^{(n)}_{i,j} ||_F^2$ is the squared Euclidean distance, and $W^{(n)}_{i,j} \doteq 2 \re{ \trace{ \mathbf{D}^{(n)}_{i,j} { \mathbf{Z}^{(n)}}^H } }$. It is easy to see that $W^{(n)}_{i,j}$ is a real Gaussian random variable
\begin{eqnarray}
W^{(n)}_{i,j} &=& \trace{  \mathbf{D}^{(n)}_{i,j} { \mathbf{Z}^{(n)}}^H +  {\mathbf{D}^{(n)}_{i,j}}^H  \mathbf{Z}^{(n)} } \nonumber \\
&=&  {\mathbf{z}^{(n)}}^H \mathbf{d}^{(n)}_{i,j} + {\mathbf{d}^{(n)}_{i,j}}^H \mathbf{z}^{(n)},
\end{eqnarray}
with
\begin{eqnarray}
\mathbb{E}[ W^{(n)}_{i,j} ] &=& 0, \\
\mathbb{E}[ {W^{(n)}_{i,j}}^2 ] &=& \mathbb{E}\left[ {\mathbf{z}^{(n)}}^H \mathbf{d}^{(n)}_{i,j} + {\mathbf{d}^{(n)}_{i,j}}^H \mathbf{z}^{(n)} \right] \nonumber \\
&=& \mathbb{E}\left[ 2 {\mathbf{z}^{(n)}}^H \mathbf{d}^{(n)}_{i,j} {\mathbf{d}^{(n)}_{i,j}}^H \mathbf{z}^{(n)} \right] \nonumber  \\
&=& 2 {d_E^{(n)}(i,j)}^2,
\end{eqnarray}
which uses the fact that  $ \mathbf{Z}^{(n)}$ has complex circularly symmetric Gaussian entries with unit variance.

The $n$-PWEP now becomes
\begin{equation}
\label{eqn:nPWEP1}
\pr{ Q_{i_1, j}^{(1)}<0, \cdots, Q_{i_n, j}^{(n)}<0  | \mathbf{H}, \mathbf{s}_{j} } = \pr{ W^{(1)}_{i_1,j} < -{d_E^{(1)}(i_1,j)}^2, \cdots, W^{(n)}_{i_n,j} < -{d_E^{(n)}(i_n,j)}^2 | \mathbf{H}, \mathbf{s}_{j} }.
\end{equation}
Notice that
\begin{equation}
\mathbf{w}^{(n)} \doteq \left( W^{(1)}_{i_1,j}, \cdots, W^{(n)}_{i_n,j} \right)^t
\end{equation}
is an $n$-dimensional real Gaussian random vector. Thus, the only remaining problem is to obtain the statistics of $\mathbf{w}^{(n)}$. It is easy to get the mean
\begin{equation}
\mathbb{E}[ \mathbf{w}^{(n)} ] = \mathbf{0}.
\end{equation}
As for the covariance matrix $\mathbb{E}[ {\mathbf{w}^{(n)}}^2 ] $, the focus is on
\begin{equation}
\mathbf{R}_{\mathbf{w}^{(n)}}(k,l) \doteq \mathbb{E}[ W^{(k)}_{i_k,j} W^{(l)}_{i_l,j} ], \forall l \geq k
\end{equation}
due to its symmetric property. The important observation is that
\begin{eqnarray}
\mathbf{z}^{(l)} &=& \vvec{\mathbf{Z}^{(l)}} \nonumber \\
&=& \vvec{\left[ \mathbf{Z}^{(k)}, \mathbf{Z}_{l-k} \right] } \nonumber \\
&=& \left[ \begin{array}{l} \mathbf{z}^{(k)} \\ \mathbf{z}_{l-k} \end{array} \right]
\end{eqnarray}
and
\begin{eqnarray}
\mathbf{d}^{(l)}_{i_l,j} &=& \mathbf{y}^{(l)}_{i_l} - \mathbf{y}^{(l)}_{j} \nonumber \\
&=& \left[ \begin{array}{l} \mathbf{y}^{(k)}_{i_l} - \mathbf{y}^{(k)}_{j} \\ \mathbf{y}_{i_l,l-k} - \mathbf{y}_{j,l-k} \end{array} \right] \nonumber \\
&=& \left[ \begin{array}{l} \mathbf{d}^{(k)}_{i_l,j}  \\ \mathbf{d}_{i_l,j,l-k} \end{array} \right].
\end{eqnarray}
Hence
\begin{eqnarray}
\mathbf{R}_{\mathbf{w}^{(n)}}(k,l) &=& \mathbb{E} \left[ \left({\mathbf{z}^{(k)}}^H \mathbf{d}^{(k)}_{i_k,j} + {\mathbf{d}^{(k)}_{i_k,j}}^H \mathbf{z}^{(k)} \right) \left({\mathbf{z}^{(l)}}^H \mathbf{d}^{(l)}_{i_l,j} + {\mathbf{d}^{(l)}_{i_l,j}}^H \mathbf{z}^{(l)} \right) \right] \nonumber \\
&=& 2 \re{ \mathbb{E} \left[{\mathbf{d}^{(k)}_{i_k,j}}^H \mathbf{z}^{(k)} \left( {\mathbf{d}^{(l)}_{i_l,j}}^H \mathbf{z}^{(l)}  \right)^* \right]  },
\end{eqnarray}
which again uses the circularly symmetric property of the i.i.d complex Gaussian random matrix $ \mathbf{Z}$. The following derivation is straightforward:
\begin{eqnarray}
\mathbb{E} \left[ {\mathbf{d}^{(k)}_{i_k,j}}^H \mathbf{z}^{(k)} \left( {\mathbf{d}^{(l)}_{i_l,j}}^H \mathbf{z}^{(l)}  \right)^* \right] &=& \mathbb{E} \left[ {\mathbf{d}^{(k)}_{i_k,j}}^H \mathbf{z}^{(k)} \left( {\mathbf{d}^{(k)}_{i_l,j}}^H \mathbf{z}^{(k)} +  {\mathbf{d}_{i_l,j,l-k}}^H \mathbf{z}_{l-k} \right)^* \right] \nonumber \\
&=&  \mathbb{E} \left[ {\mathbf{d}^{(k)}_{i_k,j}}^H \mathbf{z}^{(k)} \left( {\mathbf{d}^{(k)}_{i_l,j}}^H \mathbf{z}^{(k)} \right)^* \right] \label{eqn:R_W_1} \\
&=& {\mathbf{d}^{(k)}_{i_k,j}}^H \mathbb{E} \left[ \mathbf{z}^{(k)} {\mathbf{z}^{(k)}}^H \right] \mathbf{d}^{(k)}_{i_l,j} \nonumber \\
&=& {\mathbf{d}^{(k)}_{i_k,j}}^H \mathbf{d}^{(k)}_{i_l,j}  \nonumber \\
&=& \left\langle \mathbf{D}^{(k)}_{i_k,j}, \mathbf{D}^{(k)}_{i_l,j} \right\rangle_F
\end{eqnarray}
where for matrices $\mathbf{A}$ and $\mathbf{B}$ of the same dimension,
\begin{equation}
\left\langle \mathbf{A}, \mathbf{B} \right\rangle_F \doteq \trace{\mathbf{A}  \mathbf{B}^H}
\end{equation}
is the \emph{Frobenius inner product} \cite{HJ:85}, which induces the Frobenius norm. Equation (\ref{eqn:R_W_1}) is because $\mathbf{z}^{(k)}$ and $\mathbf{z}_{l-k}$ are independent.

Now the covariance matrix $\mathbf{R}_{\mathbf{w}^{(n)}}$ is fully characterized as
\begin{eqnarray}
\mathbf{R}_{\mathbf{w}^{(n)}}(k,l) &=& 2 \re{\left\langle \mathbf{D}^{(k)}_{i_k,j}, \mathbf{D}^{(k)}_{i_l,j} \right\rangle_F }, \forall l>k \\
\mathbf{R}_{\mathbf{w}^{(n)}}(k,k) &=& ||  \mathbf{D}^{(k)}_{i_k,j} ||_F^2, \\
\mathbf{R}_{\mathbf{w}^{(n)}}(k,l) &=&  \mathbf{R}_{\mathbf{w}^{(n)}}(l,k), \forall l<k,
\end{eqnarray}
and we conclude
\begin{equation}
\mathbf{w}^{(n)} \sim  \mathcal{N}(\mathbf{0},\mathbf{R}_{\mathbf{w}^{(n)}}).
\end{equation}
The $n$-dimensional joint Gaussian PDF of $\mathbf{w}^{(n)}$ is
\begin{equation}
f_{\mathbf{w}^{(n)}}\left(\mathbf{w}\right) = \frac{1}{(2 \pi)^{n/2} \sqrt{\det{\mathbf{R}_{\mathbf{w}^{(n)}}}} } \exp{ \left(-\frac{1}{2} \mathbf{w}^t {\mathbf{R}_{\mathbf{w}^{(n)}}}^{-1} \mathbf{w}\right)}.
\end{equation}
In the existing literature of Gaussian tail distribution, much attention has been paid to either one- or two-dimensional Q functions\cite{Simon:02}. It is clear that to analyze MIMO-HARQ with deadline $N$, one needs a general definition of the \emph{$n$-dimensional Q function} for a real Gaussian random vector $\mathbf{x} \sim \mathcal{N}(\mathbf{0},\mathbf{R}_{\mathbf{x}})$ (assuming zero mean without loss of generality)
\begin{equation}
Q_n\left( \mathbf{x}, \mathbf{R}_{\mathbf{x}} \right) \doteq  \frac{1}{(2 \pi)^{n/2} \sqrt{\det{\mathbf{R}_{\mathbf{x}}}} } \int_{\mathbf{x}(1)}^{\infty} \cdots \int_{\mathbf{x}(n)}^{\infty} \exp{ \left(-\frac{1}{2} \mathbf{w}^t {\mathbf{R}_{\mathbf{x}}}^{-1} \mathbf{w}\right)} \ud \mathbf{w}.
\end{equation}

Finally, the union bound on $P_{e}^{(n)}$ in (\ref{eqn:P_e_n_3}) becomes
\begin{equation}
P_{e}^{(n)} \leq \frac{1}{M} \sum_{j=0}^{M-1} \sum_{\substack{i_1=0 \\i_1 \neq j}}^{M-1} \cdots \sum_{\substack{i_n=0 \\i_n \neq j}}^{M-1} \mathbb{E}_{\mathbf{H}}\left[ Q_n\left( {\mathbf{d}_E^{(n)}}^2, \mathbf{R}_{\mathbf{w}^{(n)}} \right) \right],
\end{equation}
where
\begin{equation}
{\mathbf{d}_E^{(n)}}^2 \doteq \left( {d_E^{(1)}(i_1,j)}^2, \cdots, {d_E^{(n)}(i_n,j)}^2 \right)^t.
\end{equation}

\subsection{Design criterion}
\label{sec:LDC_error_crit}

The previous section studies the detection error probability after
the $n$-th ARQ round. Using the traditional union bound shifts the
focus to the $n$-PWEP. With the help of $n$-dimensional Q function
and the distribution of the decision metrics, the $n$-PWEP is
successfully calculated. However, the resulting union bound
expression is complex and does not provide insight into the optimal
LDC design. This section is devoted to developing the LDC-based HARQ
design criterion from the detection error probability point of view.
More specifically, the focus in on studying the \emph{diversity
order} of LDC-based HARQ\footnote{In this work, the diversity order
is defined as
\begin{equation*}
\textrm{$\sf{div}$} \doteq - \lim_{\textrm{$\sf{SNR}$}\rightarrow \infty} \frac{ \log P_e(\textrm{$\sf{SNR}$})}{\log \textrm{$\sf{SNR}$}}
\end{equation*}
for a \emph{fixed-rate} STC. This is the traditional definition of
diversity, which is different from the one defined in \cite{ZT:03}
and many following DMT papers.}. The main result is stated in the
following theorem.

\begin{theorem}
\label{thm:err_LDC}
The optimum diversity order for any LDC-based HARQ protocol satisfies
\begin{equation}
\label{eqn:err_LDC_proof_0}
\textrm{$\sf{div}$}_{n}^{*} = L_r  \min\{T^{(n)}, L_t\}
\end{equation}
for any ARQ round $n$, $\forall n = 1, \cdots, N$.
\end{theorem}

\emph{Proof:}
The following upper bound is straightforward:
\begin{equation}
\label{eqn:err_LDC_proof_1}
\pr{ \overline{\mathcal{A}_1}, \cdots, \overline{\mathcal{A}_{n}} } \leq \pr{ \overline{\mathcal{A}_n} }.
\end{equation}
Using this bound, the error probability (\ref{eqn:P_e_n_1}) can be bounded as
\begin{equation}
\label{eqn:err_LDC_proof_2}
P_{e}^{(n)} \leq \frac{1}{M} \sum_{j=0}^{M-1} \pr{ \mathbf{s}^{(n)} \neq \mathbf{s}_{j} | \mathbf{s}_{j} \textrm{ was sent}}.
\end{equation}
The right hand side of (\ref{eqn:err_LDC_proof_2}) is in fact the detection error probability of a STC with codeword $\mathbf{X} = \mathbf{X}^{(n)} \in \mathcal{C}^{L_t \times T^{(n)}}$. The achievable diversity order for such STC is
well-known to be $ L_r  \min\{T^{(n)}, L_t\}$. Hence
\begin{equation}
\label{eqn:err_LDC_proof_3}
\textrm{$\sf{div}$}_{n}^{*} \geq L_r  \min\{T^{(n)}, L_t\}.
\end{equation}
On the other hand, the fundamental diversity order offered by the MIMO-ARQ channel has been established in \cite{GCD:06,KS:07it}, which leads to
\begin{equation}
\label{eqn:err_LDC_proof_4}
\textrm{$\sf{div}$}_{n}^{*} \leq L_r  \min\{T^{(n)}, L_t\}.
\end{equation}
Combining (\ref{eqn:err_LDC_proof_3}) and (\ref{eqn:err_LDC_proof_4}) completes the proof.
\begin{flushright}
$\blacksquare$
\end{flushright}

Theorem~\ref{thm:err_LDC} leads to the following diversity-driven LDC design criterion for MIMO-HARQ.

\begin{crit} [Error-probability-based LDC design criterion]
\label{crit:err_LDC}
The LDC should satisfy that after the $n$-th ARQ round, $\forall n = 1, \cdots, N$, the optimum detector has a diversity order $L_r  \min\{T^{(n)}, L_t\}$.
\end{crit}

This design criterion has some practical advantages over the capacity-based Criterion~\ref{crit:cap_LDC}, as it is quite intuitive and easy to check. Meanwhile, it is also observed that those LDC-based HARQ protocols that satisfy Criterion~\ref{crit:cap_LDC} also tend to satisfy this diversity criterion.

Traditional STC design focuses on both the diversity and coding gain. A typical design would  use full diversity as a constraint and then search for good coding gain. Our effort has been only on diversity due to the following two reasons. The first is that coding gain for MIMO-HARQ is not easy to define or compute, due to the complication in the $n$-PWEP. Theoretically, each ARQ round is equivalent to a STC $\mathbf{X} = \mathbf{X}^{(n)} \in \mathcal{C}^{L_t \times T^{(n)}}$, for which the coding gain can be computed. It is unclear how to optimize these $N$ \emph{correlated} coding gains simultaneously. The second reason is that coding gain can be obtained from outer channel code as well. The diversity gain, however, has to be exploited by the proper LDC design. This is because the quasi-static channel is considered, and the only source of diversity is from the spatial domain.


\subsection{Numerical Examples and Discussions}
\label{sec:LDC_error_sim}

To gain more insight into the LDC-HARQ design based on the error
probability analysis, we return to the Alamouti example
(\ref{eqn:ala_exm}) for $N=2$. Studying this simple example leads
some interesting observations, which reveals the significance of ARQ
feedback in different settings.

A software simulation system for $\left( L_t=2, L_r=1 \right)$ MISO
i.i.d. Gaussian quasi-static fading channel with maximum ARQ rounds
$N=2$ is built. Alamouti-based protocol (\ref{eqn:ala_exm}) is
implemented, as well as the traditional repetition scheme (CC).
Maximum-likelihood decoding is performed for both ARQ rounds, which
gives the optimal decoding performance. We consider both uncoded and
coded transmission with a Gray-coded QPSK constellation on each
antenna. For the coded case, the standard 64-state rate-1/2 binary
convolutional code\footnote{The rate loss due to the termination
bits is ignored.} with octal generators (133, 171) is implemented at
the transmitter, which is followed by a standard row-in-column-out
block interleaver. Each coded packet consists of 100 symbols. The
receiver implements ML soft-output detectors with exact bit
log-likelihood ratio (LLR) computation concatenated with a bit
deinterleaver and a soft-input Viterbi algorithm  convolutional
decoder. Finally, it should be emphasized that as discussed in
\cite{SLF:08}, the random channel generation in the simulation
guarantees that the empirical channel distribution matches the true
one.

%

\begin{figure}
        \begin{center}
        \subfigure[ PER ]{
        \includegraphics[width=0.7\textwidth]{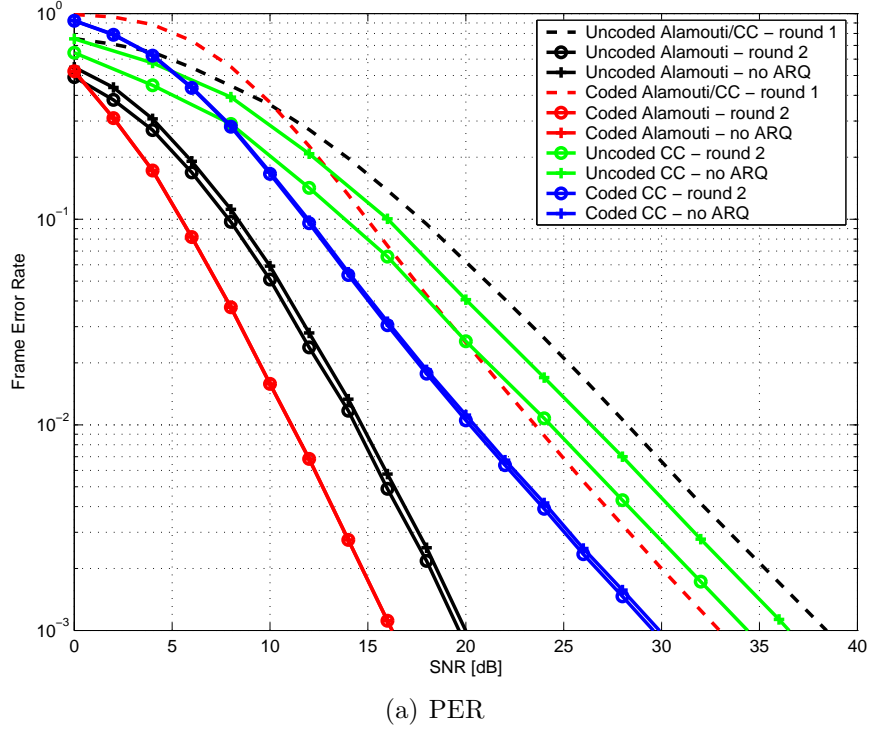}
        \label{fig:Plot9_PER} } \end{center}
        \hfil
        \subfigure[ Average Rate ]{
        \includegraphics[width=0.7\textwidth]{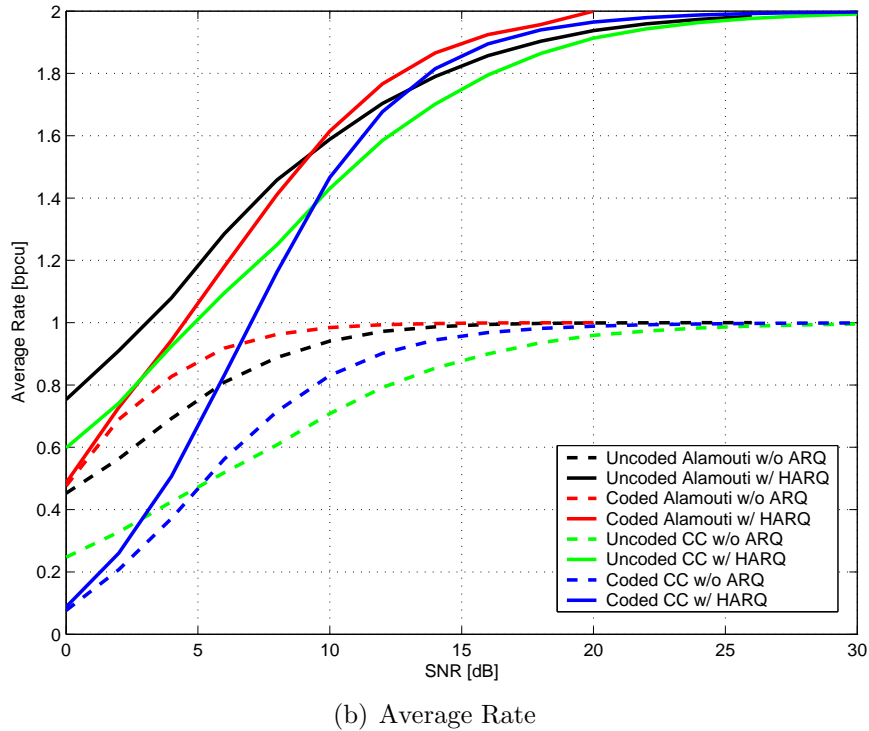}
        \label{fig:Plot9_Avg_Rate}}
    \caption{Performance comparison of two HARQ protocols (Alamouti and CC, $N=2$) and no ARQ, with both uncoded and BICM coded transmissions in a $\left( L_t=2, L_r=1 \right)$ MISO Gaussian channel. Fig.~\ref{fig:Plot9_PER}: Packet Error Rate (PER) versus SNR. Fig.~\ref{fig:Plot9_Avg_Rate}: Average Rate (bits per channel use) versus SNR.}
    \label{fig:Plot9_PER_Avg_Rate}
\end{figure}

Fig.~\ref{fig:Plot9_PER_Avg_Rate} reports both the packet error rate
and the average rate performance as a function of the average
receive SNR. As has been analytically predicted in this work, the
traditional ARQ approach which simply repeats the previous packet is
significantly inferior to the Alamouti-based protocol in all
settings. There is a clear diversity loss for the CC protocol, which
also results in the degraded average rate performance. It should be
remarked that the gains of Alamouti-based HARQ over CC are not only
in packet error rate and average rate, but also in the receiver
soft-output ML detection complexity. Alamouti detector
orthogonalizes the two transmitted symbols for independent bit LLR
generation, while the CC protocol does not render such simple
detector but an equivalent $2 \times 2$ MIMO system has to be formed
to generate bit LLR, which is of higher complexity.

In general, the gains of HARQ protocols over direct transmission without ARQ feedback are two fold:
\begin{itemize}
\item [1)] throughput;
\item [2)] decoding error probability.
\end{itemize}
The interesting observation from Fig.~\ref{fig:Plot9_PER_Avg_Rate}
is that these gains behave quite differently in different system
configurations. The average rate gain for using ARQ feedback is
remarkable for all considered protocols\footnote{Some uncoded
average rates seem to be better than coded ones in low SNR. This is
because the coded packet is longer than the uncoded one, which
counteracts the benefit from (short) channel coding.}, as can be
seen in Fig.~\ref{fig:Plot9_Avg_Rate}. However, the packet error
rate gain of the two considered HARQ protocols over no-ARQ
transmission is distinguishable for the uncoded case, but is almost
negligible for the coded one, even if the channel coding is not
capacity-approaching and the block length is  short. This can be
explained from a joint consideration of the mutual information
analysis  in Sec.~\ref{sec:LDC_MI} and the error probability
analysis in Sec.~\ref{sec:LDC_error}. Let us take $N=2$ as an
example. The decoding error probability of using ARQ is
\begin{equation}
\label{eqn:err_prob_HARQ}
P_{\textrm{$\sf{arq}$}} = \pr{ \overline{\mathcal{A}_1}, \overline{\mathcal{A}_{2}} },
\end{equation}
as compared to the no-ARQ case
\begin{equation}
\label{eqn:err_prob_noHARQ}
P_{\textrm{$\sf{no}$}\textrm{-}\textrm{$\sf{arq}$}} = \pr{\overline{\mathcal{A}_{2}} }.
\end{equation}
Hence, the gain of HARQ is
\begin{eqnarray}
P_{\textrm{$\sf{no}$}\textrm{-}\textrm{$\sf{arq}$}} - P_{\textrm{$\sf{arq}$}} &=&  \pr{\overline{\mathcal{A}_{2}} } -  \pr{ \overline{\mathcal{A}_1}, \overline{\mathcal{A}_{2}} } \nonumber \\
&=&  \pr{ \mathcal{A}_1,  \overline{\mathcal{A}_{2}} } \\
&\geq& 0 \nonumber.
\end{eqnarray}
Notice that the event $\{ \mathcal{A}_1,  \overline{\mathcal{A}_{2}}
\}$ means the receiver has a correct decoding in the first ARQ
round, but it makes an incorrect decoding as it performs decoding
based on both packets. This event is possible in the uncoded case,
as there is a good probability that the noise realization in round 1
is small (hence the decoding is correct), but very large in the
second round such that the decoding based on both transmissions
fail. Recall that the channel transfer matrix keeps constant over
ARQ rounds and hence the only source for this to happen is the noise
realization. This explains the gain in the uncoded case. However, as
the channel coding is implemented and the packet length is
increased, the impact of finite noise realizations will eventually
vanish. In fact, for the capacity-approaching coded transmission as
discussed in Sec.~\ref{sec:existingHARQ} and \ref{sec:LDC_MI}, the
noise realizations will be totally averaged out, which leads to
Equation (\ref{eqn:eventAn_subset}) and hence
\begin{equation}
\pr{ \mathcal{A}_1,  \overline{\mathcal{A}_{2}} } = 0.
\end{equation}
In summary, even with a relatively strong channel code with
reasonably long block length, the two information-theoretic
observations in Sec.~\ref{sec:existingHARQ} are generally valid, and
the ARQ retransmission leads to only an average rate gain but very
little error probability gain. On the other hand, very short coding
length makes the ARQ retranmission even more beneficial, as the
decoding error probability is improved prominently in addition to
the average rate gain.


One might notice that there is only one degree of freedom  for a
$\left( L_t=2, L_r=1 \right)$ MISO channel, but two independent
symbols are sent in the first ARQ round. This observation reveals
the essence of the HARQ gains. The concept of degrees of freedom is
from the ergodic capacity, which \emph{averages} over the channel
distribution. For a certain subset of all possible channel
realizations (together with appropriate noise realizations if the
finite noise impacts the decoding), sending two independent symbols
per channel use can result in a successful transmission. In this
case, the spectral efficiency is doubled compared to the no-ARQ
transmission, which is reflected in the average rate gain. If the
channel realization is ``bad'', the second round transmission
switches the entire protocol back to the normal STC transmission.



\section{Conclusions}
\label{sec:conc}

This paper studies the fundamental performance of Hybrid ARQ
protocols in a multiple-antenna channel. Unlike previous works that
focus on the high SNR asymptotics, a general framework to optimize
the average rate for any HARQ protocol with fixed SNR is presented.
This general result is then applied to study two well-adopted HARQ
protocols, incremental redundancy and packet combining (Chase).
Space-time coded HARQ transmission is investigated under this
framework, in which the capacity-based LDC design criterion is
derived. Several existing LDCs are evaluated with this criterion,
both analytically and numerically. A different design criterion
based on the error probability analysis of LDC-HARQ is also
presented. Numerical examples reveal different types of advantages
of LDC-HARQ, and how they vary in different settings.

There are several interesting problems that have not been addressed
in this work, which are the subjects of potential future work. For
several MIMO and ARQ configurations, the existing LDCs are proved to
be optimal in terms of average rate. However, there are some other
settings where none of the known codes approach the optimal
performance, e.g., the $(L_t=4, L_r=2, N=4)$ configuration in
Fig.~\ref{fig:Plot7_MIMO_Lt4Lr2N4}. Numerical design/search of
optimal codes (e.g., similar to the approach in \cite{HH:02}) is an
interesting research topic. Another possible direction is to
incorporate the receiver decoding complexity into the LDC design
criteria. Recall that the Alamouti-based protocol enjoys benefits of
not only optimal average rate, excellent error performance, but also
easy decoding in each ARQ rounds. Similarly, one can ask for the LDC
to not only  satisfy the design criteria presented in this work, but
also has the ``fast-decodable'' property \cite{BHV:09} for
$\mathbf{X}^{(n)}$ in each ARQ round $n$. One might have noticed
that for the LDC-based protocol to work, the channel has to remain
constant over different ARQ rounds, which is the assumption of this
paper. For time-varying channels, protocols exploiting both time and
spatial diversity should be considered, and how to efficiently
design the corresponding protocol could be another interesting
problem.

\bibliographystyle{IEEEtran}
\bibliography{ARQ}
\end{document}